%% file: ANA-STDM-2023-16-PAPER.tex
\documentclass[cernpreprint, UKenglish, texmf, orcidlogo]{atlasdoc}
 
\usepackage{atlaspackage}
\usepackage{atlasbiblatex}
 
\usepackage{atlasphysics}
 
\usepackage{graphicx}
\usepackage{grffile}
\usepackage{multirow}

\DeclareFieldFormat{labelnumberwidth}{\mkbibbrackets{#1}}
\graphicspath{{logos/}{figures/}}
 
\usepackage{ANA-STDM-2023-16-PAPER-defs}

\AtlasTitle{Measurement of vector boson production cross sections and their ratios using $pp$ collisions  at $\sqrt{s}=13.6$\,TeV with the ATLAS detector}
 
\AtlasVersion{2.5}
 
\AtlasAbstract{
Fiducial and total $W^\pm$ and $Z$ boson cross sections, their ratios and the ratio of top-antitop-quark pair and $W$-boson fiducial cross sections are measured in proton--proton collisions at a centre-of-mass energy of $\sqrt{s}=13.6$\,TeV, corresponding to an integrated luminosity of 29~fb$^{-1}$ of data collected in 2022 by the ATLAS experiment at the Large Hadron Collider. The measured fiducial cross-section values for $W^+\to \ell^+\nu$, $W^-\to \ell^-\bar{\nu}$, and $Z\to \ell^+\ell^-$ ($\ell=e$ or $\mu$) boson productions are $4250\pm 150$\,pb, $3310\pm 120$\,pb, and $744\pm 20$\,pb, respectively, where the uncertainty is the total uncertainty, including that arising from the luminosity of about 2.2\%. The measurements are in agreement with Standard-Model predictions calculated at next-to-next-to-leading-order in $\alpha_s$, next-to-next-to-leading logarithmic accuracy and next-to-leading-order electroweak accuracy.
}

\PreprintIdNumber{CERN-EP-2024-069}

\AtlasJournalRef{\PLB 854 (2024) 138725}
\AtlasDOI{10.1016/j.physletb.2024.138725}

\hypersetup{pdftitle={ATLAS document},pdfauthor={The ATLAS Collaboration}}
 
\begin{document}
 
\maketitle

\section{Introduction}
 
Precision measurements of the \Wpm\ and \Z\ cross sections and their ratios at the Large Hadron Collider (LHC) provide an excellent probe of quantum chromodynamics (QCD) and of the proton structure. Measurements of \Wpm\ and \Z\ cross sections have been performed
by the ATLAS Collaboration~\cite{PERF-2007-01} at centre-of-mass energies of 2.76\,TeV~\cite{STDM-2018-06}, 5\,TeV~\cite{HION-2018-02},
7\,TeV~\cite{STDM-2012-20}, 8\,TeV~\cite{STDM-2016-04,STDM-2017-13}, and 13\,TeV~\cite{STDM-2015-03, STDM-2018-14}
and by the CMS~\cite{CMS-CMS-00-001} Collaboration
at centre-of-mass energies of 7\,TeV~\cite{CMS-EWK-10-005,CMS-SMP-13-003}, 8\,TeV~\cite{CMS-SMP-12-011,CMS-SMP-14-003}, and 13\,TeV~\cite{CMS-SMP-18-012}.
The experimental precision of such measurements performed within the fiducial region defined by the detector acceptance has reached percent level,
with sub-percent level precision for cross-section ratios.
 
In this letter, measurements of the inclusive fiducial and total cross sections of \Wptolnu, \Wmtolnu, their ratio, and their ratios to the $Z\to \ell^+\ell^-$
cross section are presented. The measurements are made at an increased LHC centre-of-mass energy of 13.6\,TeV, using a data sample corresponding to 29~\ifb,
with $\ell$ referring to muons or electrons.
The fiducial cross section for $Z\to \ell^+\ell^-$ production in the same data sample from this analysis has been published together with the top-antitop-quark pair \tt\ production cross section
and their ratio in Ref.~\cite{TOPQ-2023-21}. Here, the first measurement of the ratios between \tt\ and \Wpm-boson fiducial cross sections in the same data sample is also presented to probe different parton densities of the proton.
 
The measurements are performed by using profile-likelihood (PLH) fits~\cite{Rolke:2004mj} to the inclusive data in the four fiducial single-lepton
channels \Wptoenu, \Wptomunu, \Wmtoenu, and \Wmtomunu, and the two fiducial dilepton channels \Ztoee\ and \Ztomm\ to extract the \W\ and \Z\ boson cross sections, and their ratios
\sWptolnu, \sWmtolnu, \sWtolnu, \sZtoll, \RWZ, and \RWpm.
Similarly, combined fits of the four \W-boson channels, the two \Z-boson channels, and the two \tt\ channels from Ref.~\cite{TOPQ-2023-21} are used to extract the cross-section ratios \RWtt, \RWptt, and \RWmtt.
 
The measurements are compared with theoretical predictions calculated at next-to-next-to-leading-order (NNLO). The predictions are supplemented by the resummation of logarithmically enhanced contributions in the low transverse-momentum region of the lepton pairs at next-to-next-to-leading-logarithmic (NNLL)
accuracy in QCD, plus next-to-leading-order (NLO) electroweak (EW) accuracy, using various state-of-the-art parton distribution functions (PDFs). The dependence of the cross sections on the centre-of-mass energy is also tested.


\section{The ATLAS detector}
 
The ATLAS experiment~\cite{PERF-2007-01,GENR-2019-02} at the LHC is a multipurpose particle detector
with a forward--backward symmetric cylindrical geometry and a near \(4\pi\) coverage in
solid angle.\footnote{\AtlasCoordFootnote}
It consists of an inner tracking detector (ID) surrounded by a thin superconducting solenoid
providing a \qty{2}{\tesla} axial magnetic field, electromagnetic and hadronic calorimeters, and a muon spectrometer.
The inner tracking detector covers the pseudorapidity range \(|\eta| < 2.5\).
It consists of silicon pixel, silicon microstrip, and transition radiation tracking detectors.
Lead/liquid-argon (LAr) sampling calorimeters provide electromagnetic (EM) energy measurements
with high granularity within the region \(|\eta|< 3.2\).
A steel/scintillator-tile hadronic calorimeter covers the central pseudorapidity range (\(|\eta| < 1.7\)).
The endcap and forward regions are instrumented with LAr calorimeters
for EM and hadronic energy measurements up to \(|\eta| = 4.9\).
The muon spectrometer surrounds the calorimeters and is based on
three large superconducting air-core toroidal magnets with eight coils each.
The field integral of the toroids ranges between \num{2.0} and \qty{6.0}{\tesla\metre}
across most of the detector.
The muon spectrometer includes a system of precision tracking chambers up to \(|\eta| = 2.7\) and fast detectors for triggering up to \(|\eta| = 2.4\).
The luminosity is measured mainly by the LUCID--2 detector~\cite{Avoni:2018} which is located close to the beampipe.
A two-level trigger system is used to select events~\cite{Aad:2886407}.
The first-level trigger is implemented in hardware and uses a subset of the detector information
to accept events at a rate below \qty{100}{\kHz}.
This is followed by a software-based trigger that
reduces the accepted event rate to \qty{3}{\kHz} on average,
depending on the data-taking conditions.
A software suite~\cite{ATL-SOFT-PUB-2021-001} is used in data simulation, in the reconstruction
and analysis of real and simulated data, in detector operations, and in the trigger and data acquisition
systems of the experiment.


\section{Data and MC samples}
\label{sec:sample}
 
The analysis is performed using data collected in 2022, the first year of the Run~3 data taking period, by the ATLAS detector in $pp$ collisions at $\sqrt{s}=13.6$\,TeV,
corresponding to an integrated luminosity of $29.0\pm 0.6$~\ifb~\cite{ATL-DAPR-PUB-2023-001}, after applying data-quality requirements~\cite{DAPR-2018-01}.
 
Monte Carlo (MC) simulated events are used to optimise the selection, determine all backgrounds except for the multi-jet process, calculate the detector acceptance factor and
to perform the final signal-extraction fit.
 
The production of \Z\ and \W\ bosons decaying to $e$, $\mu$ signal or $\tau$ background final states was simulated with the \sherpa~2.2.12 generator~\cite{Bothmann:2019yzt} using NLO matrix elements (MEs) for up to two additional partons, and leading-order (LO) ME for up to five additional partons
calculated with the \comix~\cite{Bothmann:2019yzt} and \openLoops~\cite{Buccioni:2019sur,Cascioli:2011va,Denner:2016kdg}
libraries. They were matched with the \sherpa\ parton shower (PS)~\cite{Schumann:2007mg} using the \MEPSatNLO\ prescription~\cite{Hoeche:2012yf}
and the set of tuned parameters developed by the \sherpa\ authors. The \NNPDFthreeNNLO\ PDF set~\cite{Ball:2014uwa} was used and the
samples were normalised to a NNLO prediction~\cite{Anastasiou:2003ds}.
 
Samples of diboson final states were simulated with \sherpa~2.2.12, including off-shell effects and Higgs boson contributions where appropriate.
Fully leptonic final states and semi-leptonic final states were generated using MEs at NLO accuracy in QCD for up to one additional parton and at
LO accuracy for up to three additional parton emissions.
 
The production of a top-quark pair \tt\ and the associated production of a single top-quark and a \W\ boson (\Wt) were modelled using
the \powhegbox~v2 generator~\cite{Frixione:2007nw,Nason:2004rx,Frixione:2007vw,Alioli:2010xd} interfaced to \pythia~8.307~\cite{Bierlich:2022pfr} with the parton shower tune A14~\cite{ATL-PHYS-PUB-2014-021}.
The MEs were calculated at NLO precision in QCD using the \NNPDFthreeNNLO\ PDF set. The \tt\ sample was normalised to the cross-section
prediction at NNLO in perturbative QCD including the resummation of NNLL soft-gluon terms
calculated using \TOPpp[2.0]~\cite{Beneke:2011mq,Cacciari:2011hy,Baernreuther:2012ws,Czakon:2012zr,Czakon:2012pz,Czakon:2013goa,Czakon:2011xx}.
 
The effect of multiple interactions in the same and neighbouring bunch crossings (pile-up) was modelled
by overlaying~\cite{SIMU-2020-01} the original hard-scattering event with simulated inelastic $pp$ events
generated by EPOS~2.0.1.4~\cite{Porteboeuf:2010um,Pierog:2013ria} and \pythia~8.307.
The events generated with EPOS used the EPOS LHC tune, while the \pythiae\ events used the \NNPDFtwoLO\ PDF set~\cite{Ball:2012cx}
and the A3 tune~\cite{ATL-PHYS-PUB-2016-017}. The ATLAS detector response was simulated by the \geantf\
toolkit~\cite{Agostinelli:2002hh} with the full simulation of the ATLAS detector~\cite{SOFT-2010-01}.
The simulated samples were then processed with the same software framework as the data.


\section{Object definition and event selection}
 
Events are selected if they include at least one lepton that is matched to an object identified by a single-lepton trigger. The lowest transverse momentum \pt threshold in the single-lepton triggers is $26\,\gev$ for electrons and $24~\gev$ for muons~\cite{Aad:2886407,TRIG-2018-05,TRIG-2018-01}.
Events are also required to have at least one reconstructed collision vertex with two or more associated tracks
with \pt $>$ 500~\mev. The vertex with the highest $\sum \pt^2$ of the associated tracks is taken as the primary vertex.
 
Electron candidates are reconstructed from clusters in the EM  calorimeter matched to charged tracks measured in the ID. The electron candidates are required to satisfy \textit{Tight} identification criteria~\cite{EGAM-2018-01}, validated to be still working well for data taking at 13.6\,TeV, and to have $p_\mathrm{T}>$ 27\,GeV and $|\eta|<2.47$, excluding the transition region between the barrel and endcap calorimeters, $1.37<|\eta|<1.52$.
 
Muon candidates are reconstructed from tracks in the muon spectrometer matched to tracks from the ID. The muon candidates are required to satisfy \textit{Medium} identification criteria~\cite{MUON-2018-03}, adapted for data taking at 13.6\,TeV,  with $p_\mathrm{T}>27$\,GeV and $|\eta|<2.5$.
 
Leptons must originate from the primary vertex by requiring $|z_0 \sin\theta | < 0.5$\,mm, where $z_0$ is the $z$-coordinate of the track at the point of closest approach to the $z$-axis. Furthermore, the significance of the transverse impact parameter, defined by the distance of closest approach of the track to the primary vertex point in the $r-\phi$ projection $|d_0|$, divided by its estimated uncertainty $\sigma (d_0)$, is required to satisfy $|d_0|/\sigma (d_0) <5$ for electrons and $|d_0|/\sigma (d_0) <3$ for muons.
 
To select prompt leptons, tight isolation criteria are imposed on the charged leptons. Requirements are applied on the transverse momentum sum of all ID tracks within a variable cone around the lepton, $p_\mathrm{T}^\mathrm{varcone30}$, where the maximum cone size is $\Delta R = 0.3$, shrinking for larger $p_\mathrm{T}$~\cite{EGAM-2018-01}, and on the transverse energy $E_\mathrm{T}^\mathrm{cone20}$ in a cone of $\Delta R = 0.2$ around the lepton: $p_\mathrm{T}^\mathrm{varcone30}/p_\mathrm{T}<0.06$ and $E_\mathrm{T}^\mathrm{cone20}/p_\mathrm{T}<0.06$ for electrons and $p_\mathrm{T}^\mathrm{varcone30}/p_\mathrm{T}<0.04$ and $E_\mathrm{T}^\mathrm{cone20}/p_\mathrm{T}<0.15$ for muons.
 
Jets are used for the calculation of the missing transverse momentum (see below). They are reconstructed using a particle flow algorithm~\cite{PERF-2015-09} that exploits both calorimeter and ID informations.
The \antikt\ algorithm~\cite{Cacciari:2008gp,Fastjet} with a radius parameter $R=0.4$ is used.
The jet energy calibration is based on Run~3 simulation and \emph{in-situ} calibration~\cite{JETM-2018-05}, adapted for data taking at 13.6\,TeV.
Calibrated jet candidates are required to have $\pt > 20~\gev$ for  $|\eta| < 2.5$ and $\pt > 30~\gev$ for $ 2.5 <  |\eta| < 4.5$.
To suppress jets originating from pile-up, jets with \pt\ below 60~\gev are required to satisfy a neural-network-based jet vertex tagger (NNJVT) discriminant, a successor of the jet vertex tagger  algorithm~\cite{PERF-2014-03} used during the Run 2 data taking period in the years 2015 to 2018.
 
To remove the ambiguity when one physical object is reconstructed as multiple objects, the following algorithm is used, with each operation applied in the given order. If any electron candidate is found sharing a track with any other electron candidates, the candidate with smaller $\pt$ is removed. Any electron candidate found to share a track with a muon is removed. The closest jet found within a $\Delta R$ of 0.2 of an electron candidate is removed. Any electron candidate subsequently found within $\Delta R$ of 0.4 of a jet is removed. Jets with less than three tracks associated to it found within $\Delta R$ of 0.2 or with a muon inner-detector track ghost-associated to it are removed, then muon candidates found within $\Delta R$ of 0.4 of a jet are removed.
 
The missing transverse momentum $\vec{p}_\mathrm{T}^\mathrm{miss}$, with magnitude $E_\mathrm{T}^\mathrm{miss}$, is defined as the negative of the sum of the transverse momenta of the reconstructed and calibrated physical objects, and a soft term built from all tracks that are associated with the primary vertex, excluding those used in the physics objects, is also included~\cite{ATLAS-CONF-2018-023}.
 
The object reconstruction for the $t\bar{t}$ analysis is described in detail in Ref.~\cite{TOPQ-2023-21}. In addition to the objects defined above, jets containing $b$-hadrons are identified, using the 77\% efficiency working point of the DL1d $b$-tagging algorithm~\cite{FTAG-2019-07,ATL-PHYS-PUB-2020-014}.
 
The $Z$-boson decays into a pair of electrons or muons with opposite electric charge, thus two dilepton channels, $ee$ and $\mu\mu$, are defined. In the case of the $W$-boson, the ones decaying into one lepton and a neutrino are considered, so four single lepton channels are defined according to the lepton flavour and electric charge: $e^-\bar{\nu}$, $e^+\nu$, $\mu^-\bar{\nu}$, and $\mu^+\nu$.
 
Events in the dilepton channels are required to have exactly two same-flavour leptons (electrons or muons) with opposite electric charge.
The invariant mass of the dilepton pairs is required to be in the mass range $66 < \mll < 116\,\gev$.
In the single-lepton channels, events are required to have exactly one identified and isolated lepton, $E_\mathrm{T}^\mathrm{miss}$
greater than 25\,\gev, and a transverse mass $m_\mathrm{T}^W=\sqrt{2p_\mathrm{T}^\mathrm{miss} p_\mathrm{T}^\ell (1-\cos\Delta\phi_{\ell\nu})}$ greater than 50\,\gev,
where $\Delta\phi_{\ell\nu}$ is the azimuthal angle between the charged lepton transverse momentum $\vec{p}_\mathrm{T}^{\ell}$ and $\vec{p}_\mathrm{T}^\mathrm{miss}$.
 
The event selections for the $W$ and $Z$ boson analysis are summarised in Table~\ref{tab:reco_event_selection}. The $Z\to \ell\ell$ selection is the same as in Ref.~\cite{TOPQ-2023-21} that also describes the event selection for the $t\bar{t}$ analysis. Selected \tt\ candidates are required to have exactly one electron and one muon of opposite charge. No dilepton mass, \met\ or \mtw\ requirements are imposed but the events are required to have one or two $b$-tagged jets.
 
\begin{table}[tb]
\centering
\caption{Summary of the event selection requirements.}
\label{tab:reco_event_selection}

\begin{tabular}{l|c}
\hline
Electron selections  &
\pt   $> 27$\,GeV                                                         \\
& $\aeta < 2.47$ and veto of $1.37 < \aeta < 1.52$ \\
\hline
Muon selections      &
\pt    $>$ 27\,GeV                                                        \\
& $\aeta  < 2.5$                                     \\
\hline
$W$-boson selections &
Exactly one lepton                                                        \\
& $\met > 25$\,GeV                                  \\
& $\mt\  > 50$\,GeV                                  \\
\hline
$Z$-boson selections &
Exactly two same flavour opposite charged leptons                         \\
& $66 < \mll <116$\,GeV                      \\
\hline
\end{tabular}


\end{table}


\section{Background estimation and yields}
\label{sec:background}
 
Background contributions to the \W\ and \Z\ boson final states correspond to two main categories: EW and top-quark processes, estimated by using MC simulations, and the multi-jet (MJ) background, estimated using a data-driven technique.
 
The EW background contributions include single-boson and diboson productions.
The single-boson productions correspond to \Wtotnu\ and \Ztott\  for both channels, where the subsequent leptonic decays of the $\tau$ leptons are also treated as background, \Ztoee\ and \Ztomm\ for the \W\ channels and \Wtoenu\ and \Wtomunu\ for the \Z-boson channels.
Diboson productions, \WW, \WZ, and  \ZZ\, can have similar signatures to \W\ or \Z\ events if one of the bosons decays hadronically or
invisibly or if leptonic decay products fail to satisfy the object selection. Likewise, \tt\ pair and \Wt\ production can result in
similar signatures to \W\ or \Z\ events, if one or two  \W\ bosons decay leptonically. For \W-boson selections, the contribution
of both the diboson and top-quark events to the background is small due to the lower production cross section.  For \Z-boson selections,
all EW background contributions are at sub-percent level.
 
The MJ background stems from QCD jet production, where particles within the jets are mistakenly identified as prompt isolated leptons.
It has major contributions from collimated charged and neutral pions and from non-prompt real leptons produced e.g.\ in semi-leptonic
decays of heavy quarks or in-flight pion decays.
Although this type of background processes is efficiently rejected by the isolation selection and, in the \W-boson case, by the lepton $p_\mathrm{T}$, \met, and \mtw\ requirements, MJ processes still dominate
the background in \W-boson measurements at high pile-up due to the large production cross section and large \met\ generated through jet energy mis-measurements
in the event, contributing to around 3$-$7\% of data events.
Because of the difficulties in the precise simulation of these processes, data-driven techniques are used for the estimation
of the MJ background~\cite{STDM-2015-03}.
 
The MJ background in the four \W-boson channels is estimated by performing PLH fits to the data in a fitting region enriched with MJ contribution,
where the signal and background kinematic distributions have sufficient discriminating power. The discriminating variables used in these fits are \met\ and \mtw.
The signal region (SR) is defined by the full event selection, in particular, \met$>25$\,\gev and \mtw$>50$\,\gev.
The fitting region (FR) is defined similarly to the SR except that \met$<25$\,\gev and \mtw$<50$\,\gev are required.
The signal, EW, and top-quark background contributions in this region are estimated by using MC simulations, while the MJ distributions are derived in a control region (CR1) with similar kinematic selection as the FR,
but where the lepton is required to fail to satisfy the track isolation requirement, denoted as anti-isolated.
Similarly, another control region (CR2) is defined where the lepton is also required to fail isolation but has the same kinematic selection as the SR.
The contamination from signal and other backgrounds in the CR1 and CR2 is estimated by using MC simulations and subtracted from the data to obtain the MJ templates in these regions.
Pre-scaled supporting triggers with looser or without isolation requirements corresponding to a smaller integrated luminosity than the nominal one are used to populate sufficiently the CR1 and CR2. These triggers have similar kinematic requirements to the nominal ones used to select events in the SR.
 
Several MJ templates are created for each $W$-boson decay channel by slicing the CR1 and CR2 as explained in the following.
Four mutually exclusive isolation slices are defined by varying the track isolation variable progressively further from the signal region, while the calorimeter isolation
is kept at the nominal value. The slices in the track isolation variable are equal in width and have an upper limit corresponding to 0.3 and 0.2 in the electron and muon channels, respectively.
For each isolation slice, the normalisation of the corresponding MJ template in the FR is extracted using a PLH fit to the measured \met\ or \mtw\
distribution. A transfer factor, calculated using the ratio of the MJ yields in the CR1 and CR2, is used to obtain the SR MJ yield from the FR fit result.
This procedure leads to two sets of four SR MJ yields (two fit variables and four isolation slices) for each of the four \W-boson channels.
To reduce the bias due to the track isolation on the MJ yield in the SR, the MJ estimates obtained from each isolation slice are used to build an extrapolation to the track isolation selection used in the SR.
 
Figure~\ref{fig:MJfits_method3_quadratic} shows the linear (dashed line) and quadratic (solid line) fits to the relative MJ yields as a function of the track isolation slice,
where the average value of the track isolation of the events in the respective isolation slice is used as the central value.
For each point, the uncertainty of the MJ fit in the FR is propagated into the MJ yield and enters the uncertainty of the final extrapolation.
The quadratic fits are observed to have better performance than the linear fits in each channel, verified with a $\chi^2$ criterion.
The combined mean of the two quadratic extrapolations using the \met\ or \mtw\ distribution is used as the central value for the final MJ yield in each channel, denoted $f_\mathrm{MJ}^\mathrm{SR}$ in the figure.
The first contribution to the uncertainty in the mean is calculated as the sum in quadrature between the combined error from the \mtw\ and \met\ fits, and the difference between the \mtw\ and \met\ fit results.
The second contribution to the systematic uncertainty is derived from the difference between the linear and quadratic fit results, which constitutes the dominant contribution.
 
\begin{figure}[tbp]
\centering
\subfloat[]{
\includegraphics[width=0.45\textwidth]{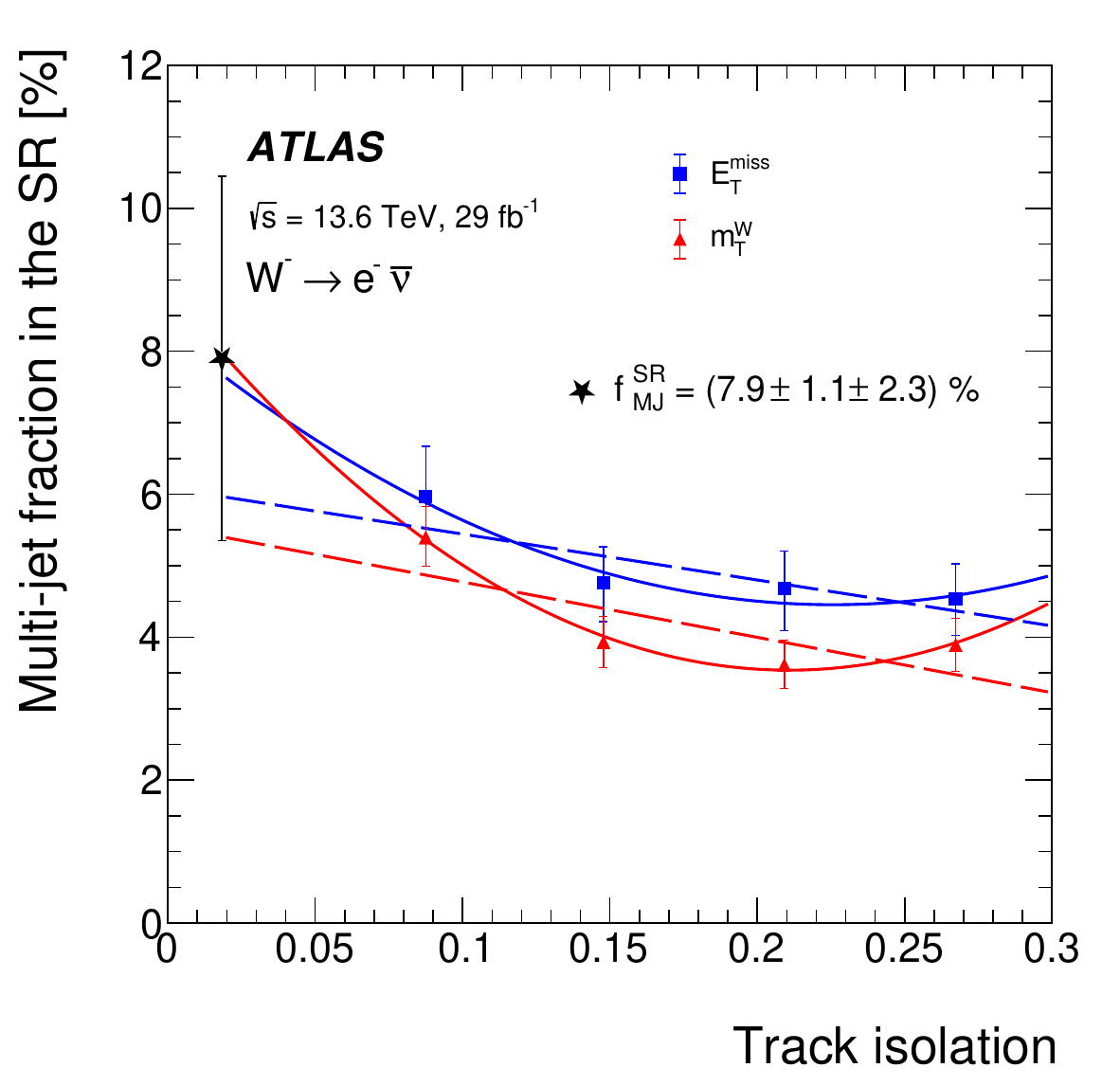}
}
\subfloat[]{
\includegraphics[width=0.45\textwidth]{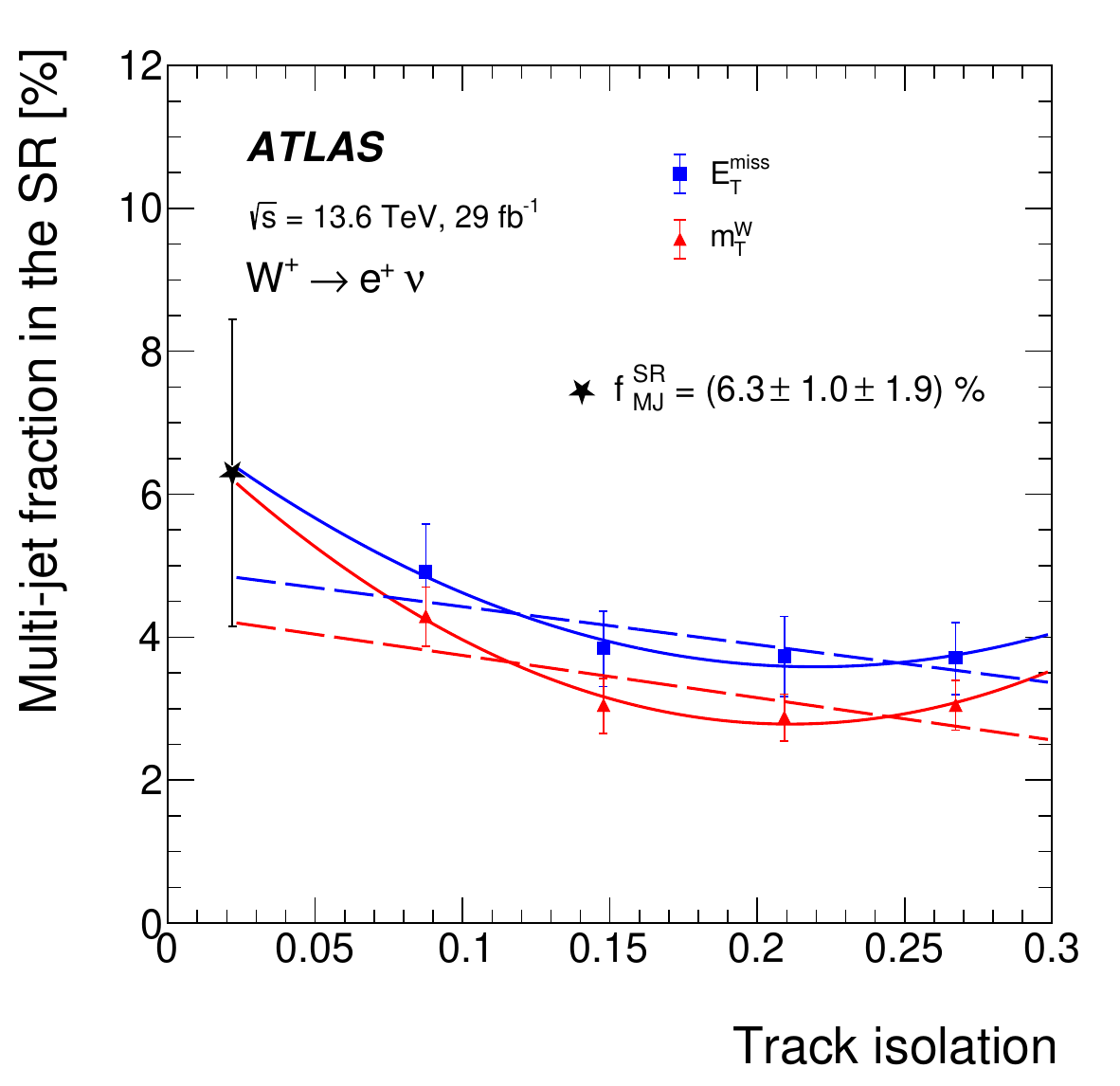}
}\\
\subfloat[]{
\includegraphics[width=0.45\textwidth]{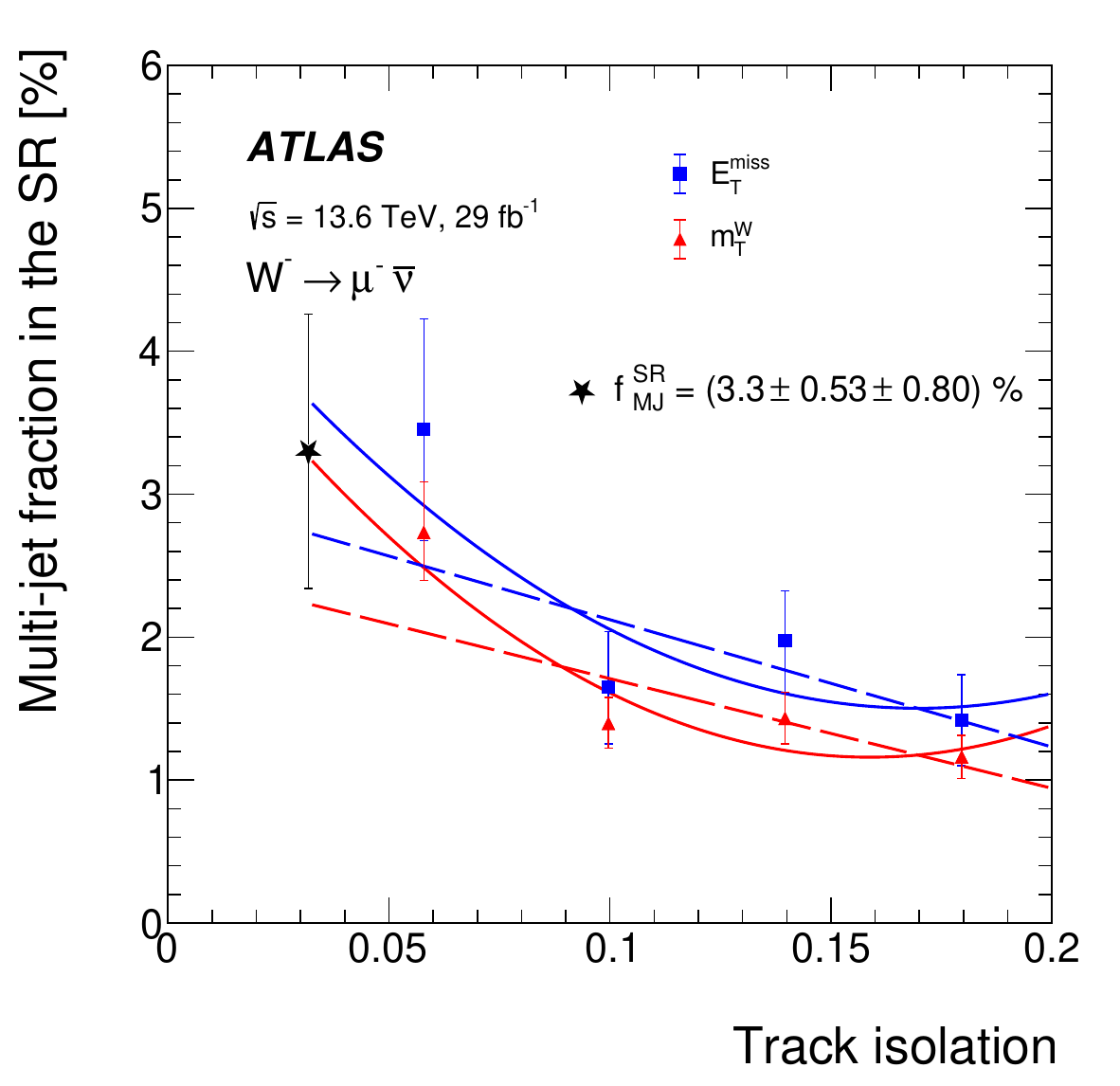}
}
\subfloat[]{
\includegraphics[width=0.45\textwidth]{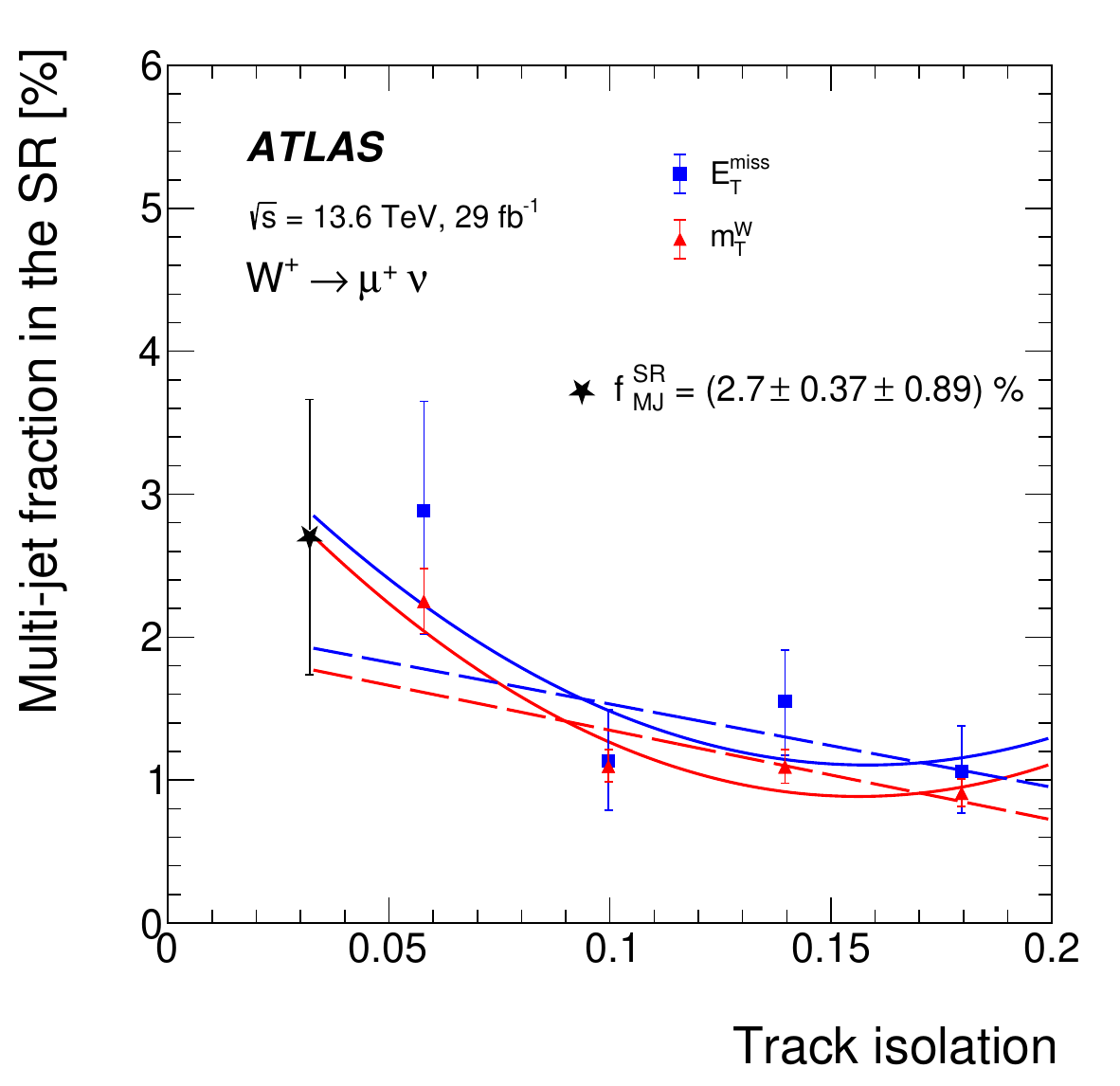}
}
\caption{Relative multi-jet yield in the SR as a function of the track isolation variable for the (a) $W^-\to e^-\bar{\nu}$, (b) $W^+\to e^+\nu$, (c) $W^-\to\mu^-\bar{\nu}$, and (d) $W^+\to \mu^+\nu$ channels, shown with squares and triangles, respectively for the $E_\mathrm{T}^\mathrm{miss}$ and $m_\mathrm{T}^W$ input measurements.
The curves represent the extrapolation of the points to the SR using a quadratic function (solid curves) or a linear function (dashed lines).
The $x$-axis corresponds to the position of the average of a track-isolation slice in the track isolation.
The star represents the final MJ fraction ($f_\mathrm{MJ}^\mathrm{SR}$) and is calculated using the combined average of the quadratic fits in each channel.
The first uncertainty corresponds to the combined uncertainty including the difference between the $m_\mathrm{T}^W$ and $E_\mathrm{T}^\mathrm{miss}$ quadratic fits,
while the second uncertainty is due to the difference between linear and quadratic fit results.}
\label{fig:MJfits_method3_quadratic}
\end{figure}

In the \Z-boson channels, a conservative upper limit on the MJ background is estimated from the number of charge misidentified leptons, calculated using the \mll\ sidebands from a same-sign lepton selection.
In the case of the electron channel, this is found to be of sub-percent level, while the contribution in the muon channel is found to be even smaller. The systematic uncertainties are also negligible. The MJ contribution in the \Z-boson channels is hence not considered in this analysis.
 
The event yields of background processes together with those from signal processes are shown in Table~\ref{tab:event_yields}. Figure~\ref{fig:dist_after_selection} shows the \mtw\ distributions in data and predictions for the four \W-boson channels. The data agrees with the prediction within the uncertainties indicated by the hashed band. The \mll\ distributions for the $Z\to \ell\ell$ analysis are presented in Ref.~\cite{TOPQ-2023-21}.
 
\begin{table}[tb]
\centering
\caption{Event yields of data and predictions after selections. Only MC statistical uncertainties are shown for the EW and top-quark processes. For the multi-jet process, the normalisation uncertainty is displayed instead. Rounding has been applied to all the yields except for those of the data.}
\label{tab:event_yields}

\scalebox{0.8}{
\begin{tabular}{l|c|c|c|c}
\hline
& $W^{-}\to e^{-}\bar{\nu}$ & $W^{+}\to e^{+}\nu$      & $W^{-}\to \mu^{-}\bar{\nu}$ & $W^{+}\to \mu^{+}\nu$   \\
\hline
$W\to e\nu$           & $43,650,000 \pm 70,000$   & $55,370,000 \pm 80,000$  & $-$                         & $-$                     \\
$W\to \mu\nu$         & $-$                       & $-$                      & $57,760,000 \pm 80,000$     & $74,900,000 \pm 90,000$ \\
$W\to \tau\nu$        & $684,000 \pm 8,000$       & $819,000 \pm 9,000$      & $906,000 \pm 10,000$        & $1,120,000 \pm 10,000$  \\
$Z\to ee/\mu\mu$      & $1,416,000 \pm 4,000$     & $1,459,000 \pm 4,000$    & $4,638,000 \pm 8,000$       & $4,903,000 \pm 8,000$   \\
$Z\to \tau\tau$       & $88,000 \pm 1,000$        & $91,000 \pm 1,000$       & $107,000 \pm 1,000$         & $111,000 \pm 1,000$     \\
$t\bar{t}$+single-top & $863,800 \pm 400$         & $905,400 \pm 500$        & $802,200 \pm 400$           & $843,400 \pm 400$       \\
$VV$                  & $93,500 \pm 300$          & $97,600 \pm 300$         & $98,500 \pm 300$            & $102,800 \pm 300$       \\
Multi-jet             & $4,000,000\pm1,000,000$   & $4,000,000\pm1,000,000$  & $2,100,000\pm700,000$       & $2,200,000\pm800,000$   \\
\hline
Total predicted                & $51,000,000\pm1,000,000$  & $63,000,000\pm1,000,000$ & $66,400,000\pm700,000$      & $84,200,000\pm800,000$  \\
\hline
Data                  & $50,748,537$              & $62,610,338$             & $65,053,470$                & $82,360,980$            \\
\hline
\end{tabular}
}
 
\scalebox{0.8}{
\begin{tabular}{l|c|c}
\hline
& $Z\to e^{+}e^{-}$      & $Z\to \mu^{+}\mu^{-}$   \\
\hline
$Z\to ee$             & $7,710,000 \pm 10,000$ & $-$                     \\
$Z\to \mu\mu$         & $-$                    & $13,961,000 \pm 10,000$ \\
$Z\to \tau\tau$       & $2,500 \pm 200$        & $3,900 \pm 200$         \\
$W\to \ell\nu$        & $139 \pm 57$           & $40 \pm 27$             \\
$t\bar{t}$+single-top & $24,950 \pm 50$        & $37,000 \pm 50$         \\
$VV$                  & $14,960 \pm 80$        & $23,800 \pm 100$        \\
\hline
Total predicted                & $7,750,000 \pm 10,000$ & $14,030,000 \pm 10,000$ \\
\hline
Data                  & $7,812,978$            & $14,242,875$            \\
\hline
\end{tabular}
}


\end{table}
 
\begin{figure}[!htb]
\centering
\subfloat[]{
\includegraphics[width=0.495\textwidth]{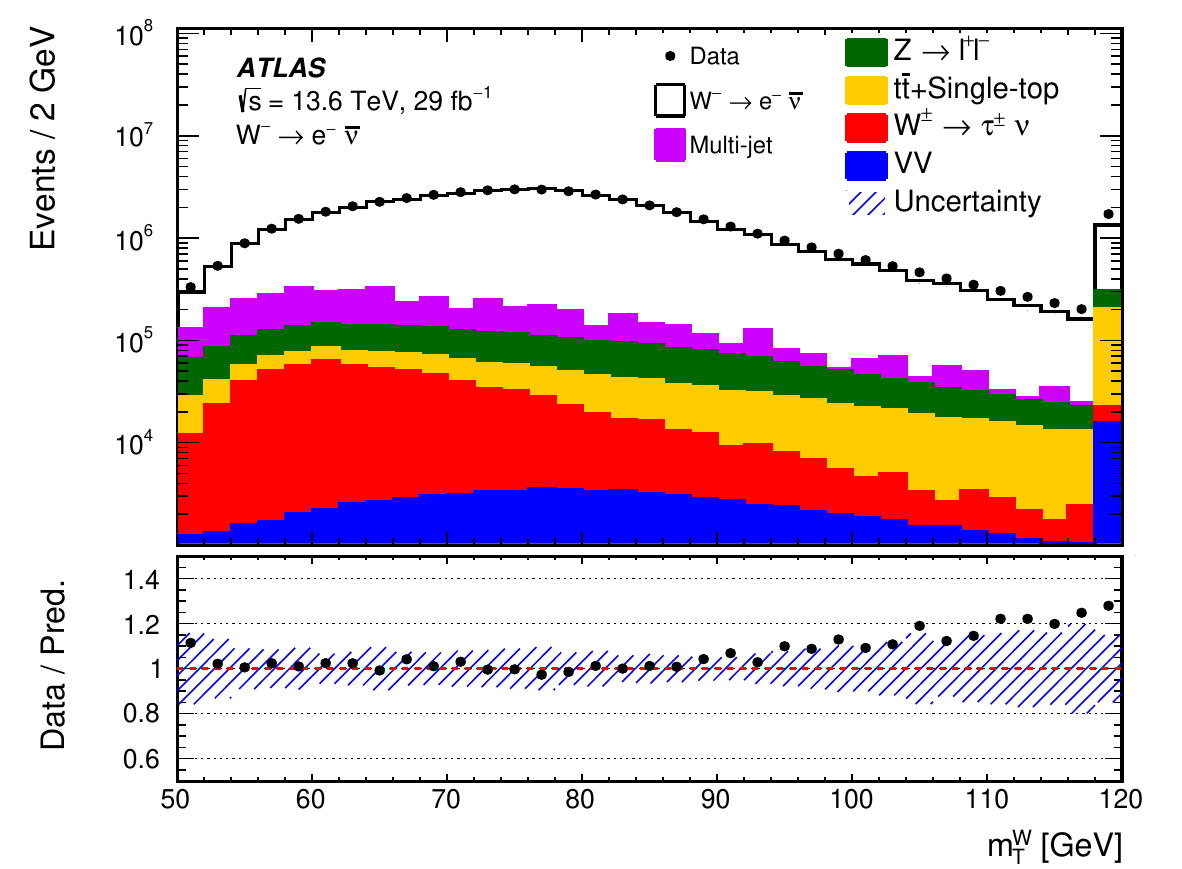}
}
\subfloat[]{
\includegraphics[width=0.495\textwidth]{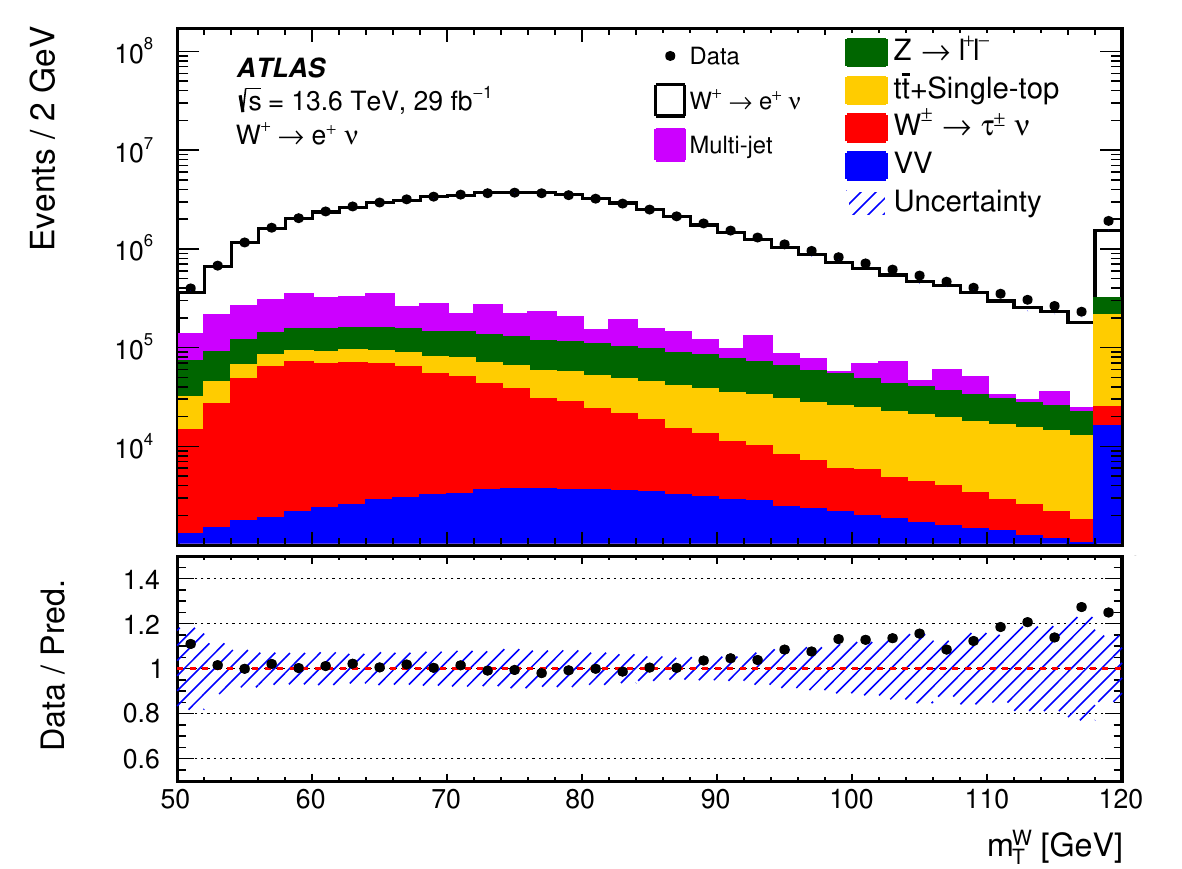}
}\\
\subfloat[]{
\includegraphics[width=0.495\textwidth]{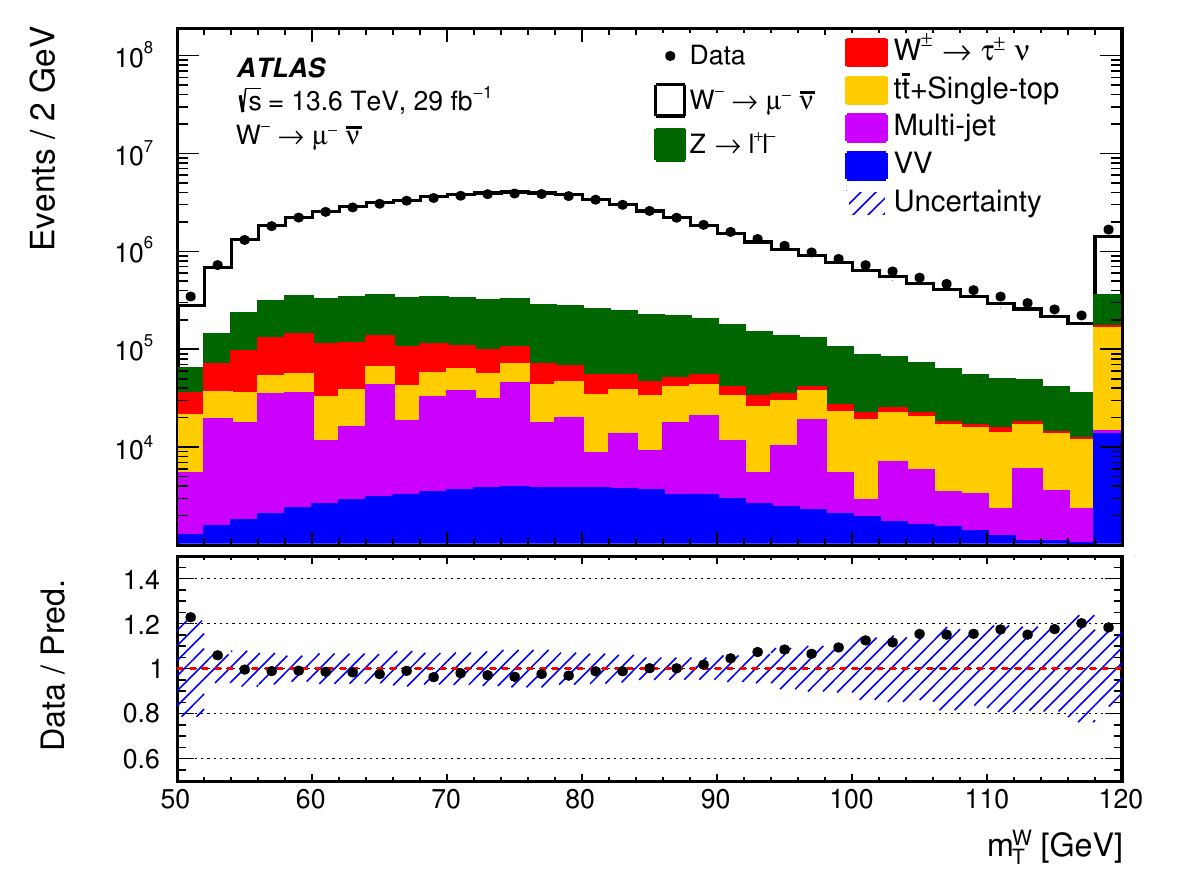}
}
\subfloat[]{
\includegraphics[width=0.495\textwidth]{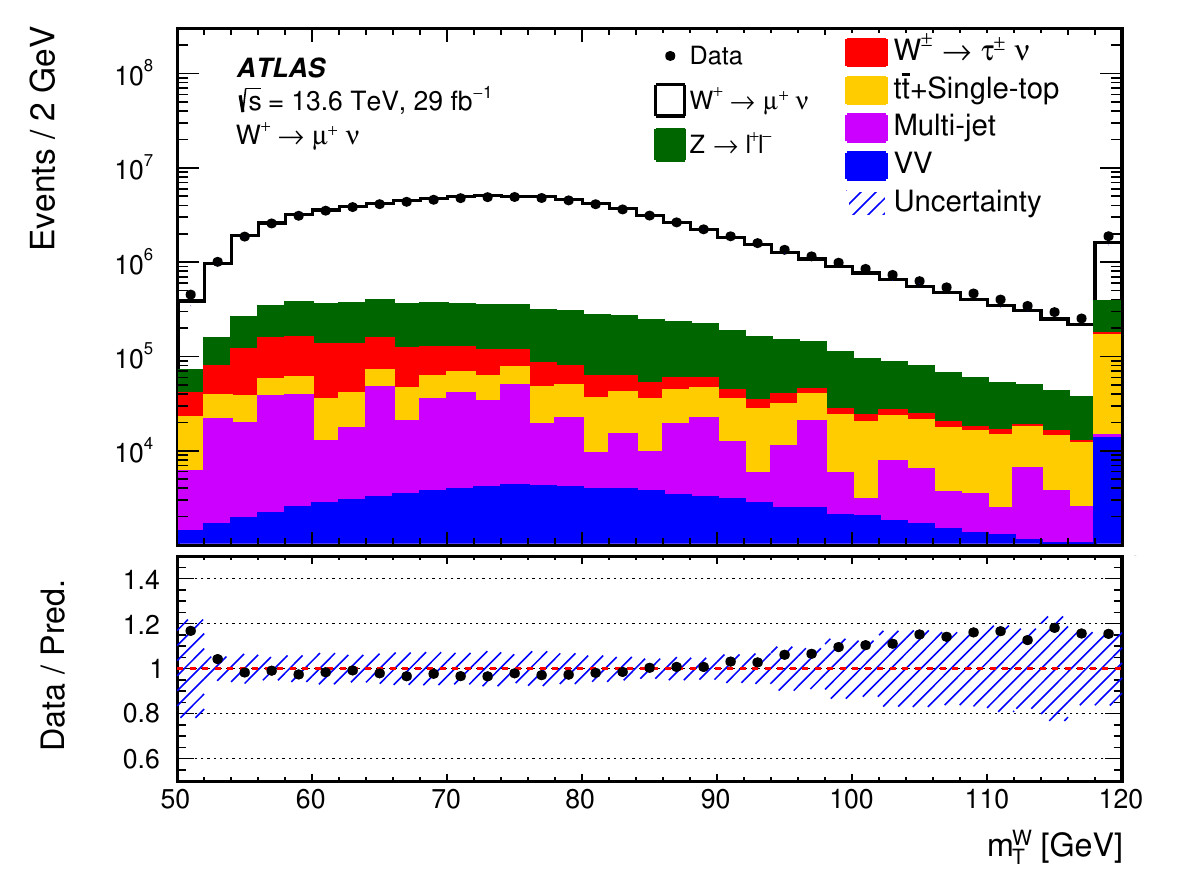}
}
\caption{Comparison of data (dots) and predictions (histograms) for the $m_\mathrm{T}^{W}$ distributions in the (a) $W^-\to e^-\bar{\nu}$, (b) $W^+\to e^+\nu$, (c) $W^-\to\mu^-\bar{\nu}$, and (d) $W^+\to\mu^+\nu$ channels. The hashed band in the ratio plot denotes the total systematic uncertainty in the prediction. The rightmost bins contain the overflow events.}
\label{fig:dist_after_selection}
\end{figure}


\section{Systematic uncertainties}
\label{sec:unc}
 
The uncertainty of the integrated luminosity is 2.2\%, based on the LUCID--2 detector~\cite{Avoni:2018} estimate, which is used for the primary luminosity measurements.
For the electron efficiency corrections~\cite{EGAM-2018-01}, extra systematic uncertainties are considered by comparing the simulations between Run 2 and Run 3.
The uncertainty of the muon efficiency corrections is obtained using the Run 3 data sample and simulations by independently varying the inner detector and muon spectrometer components, as determined from \Ztomm\ events, following Ref.~\cite{MUON-2018-03}.
The electron energy calibration uncertainty is derived using Run 2 data sample and simulations, as described in Ref.~\cite{EGAM-2018-01}, with dedicated uncertainties covering the difference between Run 2 and Run 3 simulations. The muon energy calibration uncertainty is obtained using Run 3 data, based on the approach described in Ref.~\cite{MUON-2018-03}.
The lepton trigger uncertainties are obtained by comparing the trigger efficiency results of Run 3 data with simulations.
The jet energy scale and resolution uncertainties are estimated by using the Run 2 data sample, as described in Ref.~\cite{JETM-2018-05}, and an additional uncertainty is used to cover the difference between Run 2 and Run 3 by comparing simulations, as described in Ref.~\cite{TOPQ-2023-21}. A conservative 10\% uncertainty per jet is considered to cover the differences between data and simulations in the measured NNJVT efficiencies.
Besides the propagation of systematic uncertainties of all other physics objects, those of the track soft term are also considered for \met~\cite{ATLAS-CONF-2018-023}.
The uncertainty in the pile-up is determined by varying the average number of interactions per bunch-crossing by 4\% in the simulation and it is accounted for with dedicated studies.
The systematic uncertainties arising from normalisation of the multi-jet background are described in Section~\ref{sec:background}.
 
The theoretical uncertainties are categorised into two distinct components: the uncertainty $\delta A$ on the \textit{acceptance} $A$,
the ratio of the number of events in the fiducial volume and the total phase space at the particle level, and the uncertainty $\delta C$  on the
\textit{correction factor}  $C$, the ratio of the expected signal events at the detector level and at the particle level in the fiducial volume, defined by lepton $p_\mathrm{T}^\ell>27$\,\gev, lepton $|\eta^\ell|<2.5$, \met$>25$\,\gev and \mtw$>50$\,\gev for the $W$-boson processes and the same $p_\mathrm{T}^\ell$ and $|\eta^\ell|$ selections and $66<m_{\ell\ell}<116$\,\gev for the $Z$-boson processes.
For each theoretical uncertainty, $\delta A$  is propagated into the measurement of the total cross section, whereas $\delta C$
is included as a nuisance parameter in the PLH fit used to extract the fiducial cross section.
 
Various sources of the theoretical uncertainty are evaluated using the \textsc{Sherpa} signal samples.
The QCD scale uncertainty is defined by the symmetrised envelope of seven-point variations of the renormalisation and factorisation scales, corresponding to varying $\mu_{R}$  and $\mu_{F}$ independently by factors of 1/2 and 2 to the combinations of $(\mu_{R}, \mu_{F}) = (\mu_{R}/2, \mu_{F}/2), (2\mu_{R}, 2\mu_{F}), (\mu_{R}, 2\mu_{F}), (2\mu_{R}, \mu_{F}), (\mu_{R}, \mu_{F}/2)$, and $(\mu_{R}/2, \mu_{F})$.
The PDF uncertainty is estimated based on the internal replicas of the \NNPDFthreeNNLO\ set, which enter the fit as individual nuisance parameters.
The PDF choice uncertainty is estimated by comparing the predictions calculated with \NNPDFthreeNNLO\ and with PDF4LHC21~\cite{Ball:2022hsh}.
The $\alpha_s$ uncertainty is calculated by comparing two different $\alpha_s$ values used in PDF4LHC21.
 
The contribution of each of the background processes based on simulated samples is at percent level or smaller. A conservative modelling uncertainty of 5\% covering the PDF input, $\alpha_\mathrm{s}$, and QCD scale variations, is assigned for background processes from the \W\ and \Z\ bosons and the theoretical uncertainty is 10\% for the diboson processes. The modelling uncertainties are 5.1\% and 3.5\% for the $t\bar{t}$ and single-top processes, respectively, as evaluated in Ref.~\cite{TOPQ-2023-21}.
 
The results for the $Z$ boson production are derived from fits to the two \Z-boson channels as reported in Ref.~\cite{TOPQ-2023-21}.
The other results are derived from simultaneous fits of the four \W-boson and the two \Z-boson channels.
The lepton efficiency corrections, trigger efficiency and energy calibration uncertainties are treated as uncorrelated between electron and muon channels. The multi-jet background uncertainty is also considered to be uncorrelated. Other experimental uncertainties (uncertainty sources related to jet, \met\ , pile-up and luminosity) are considered to be fully correlated.
The modelling uncertainties of electroweak and top-quark processes are regarded as fully uncorrelated. The uncertainties in signal modelling are fully correlated when combining electron and muon channels for the same boson. However, when combining channels of different bosons, these uncertainties are treated as uncorrelated sources due to the different production modes at the LHC.
 
The dominant uncertainty sources are channel dependent. The \Z-boson cross sections are limited by the luminosity and lepton correction uncertainties. For the \W-boson cross sections, the multi-jet background and jet-related uncertainties play important roles. However, for $W^+/W^-$-boson cross-section ratios, the dominant experimental uncertainty sources largely cancel out. The multi-jet background uncertainties become dominant sources for $W^+/W^-$-boson cross-section ratios since the multi-jet $W^+$ and $W^{-}$ boson uncertainties in each lepton channel are regarded as independent sources. Similarly,  in the ratio of \Wpm-boson and \Z-boson cross sections, the jet and multi-jet background uncertainties only affect the \W-boson cross section but have no impact on the \Z-boson cross section so they become dominant sources. In terms of the ratio of the \tt\ and \W\ fiducial cross section, the \tt\ modelling and multi-jet background uncertainties are significant as they do not cancel out and, additionally, uncertainties stemming from jet and lepton trigger efficiency, as well as background modelling except the multi-jet, also do not fully cancel out and so they also play important roles.


\section{Results}
 
The cross sections \sWptolnu, \sWmtolnu, \sWtolnu, and \sZtoll\
are extracted from PLH fits to the inclusive data in the four fiducial single-lepton
channels \Wptoenu, \Wptomunu, \Wmtoenu, and \Wmtomunu, and the two fiducial dilepton channels \Ztoee\ and \Ztomm.
The statistical model of the PLH fits is constructed in the following form:
\begin{equation}
L(\vec{n};\mu_s,\vec{\theta})=\prod_{c\in \mathrm{channels}}\mathrm{Pois}(n_{\mathrm{data}}|\mu_{s,c} S_{c}(\vec{\theta}) + B_{c}(\vec{\theta}))\prod_{i\in \mathrm{NPs}}G(\theta_i)\,,
\label{eq:likelihood_formula}
\end{equation}
where  $\mu_s$ is the signal strength, which represents the ratio of the measured signal cross section over the predicted value, $S_{c}$ is the expected number of signal events, and $B_{c}$ is the expected background. The normalisation of the background contributions is determined by the fit and constrained by the background cross-section uncertainties. The quantity $\theta_i$ is a nuisance parameter (see Section~\ref{sec:unc})  which is constrained by a Gaussian term $G(\theta_i)$. The probability model is the product of the Poisson distributions (``Pois'') in each channel. The theoretical uncertainties of the \W- and \Z-boson signal processes are incorporated into the fitting by introducing normalisation factors on the signal samples. The normalisation factors are calculated by comparing the correction factors of the nominal and theoretical variations, and taking the relative difference.
 
The ratios of the fiducial cross sections are extracted using fits as well. For the $W^+/W^-$-boson cross-section ratio, \RWpm, the \Wp-boson signal strength $\mu_{W^+}$ is expressed as the product $\RWpm  \mu_{W^-}$  in Eq.~(\ref{eq:likelihood_formula}), thus the likelihood formula for the ratio is written as:
\begin{align*}
L(\vec{n};\mu_s,\vec{\theta}) =&\prod_{c\in W^+\ \mathrm{channels}}\mathrm{Pois}(n_{\mathrm{data}}|\RWpm\ \mu_{W^-} S_{c}(\vec{\theta}) + B_{c}(\vec{\theta}))\times \\
&\prod_{c\in W^-\ \mathrm{channels}}\mathrm{Pois}(n_{\mathrm{data}}|\mu_{W^-} S_{c}(\vec{\theta}) + B_{c}(\vec{\theta}))\times \\
&\prod_{c\in Z\ \mathrm{channels}}\mathrm{Pois}(n_{\mathrm{data}}|\mu_{Z} S_{c}(\vec{\theta}) + B_{c}(\vec{\theta})) \prod_{i\in \mathrm{NPs}}G(\theta_i)\,.
\end{align*}
The ratios \RWZ, \RWtt, \RWptt, and \RWmtt\ are extracted in the same manner.
In the latter cases, the \tt\ inputs and uncertainties are obtained from the \tt\ cross-section analysis at 13.6\,TeV~\cite{TOPQ-2023-21}.

Fiducial cross sections, $\sigma^\mathrm{fid}$, are calculated by multiplying the signal-strength parameter $\mu$ with the nominal predicted fiducial cross section
calculated with the \SHERPA\ signal samples (see Section~\ref{sec:sample}).
Uncertainties in the signal-strength parameter $\mu$ are propagated into $\sigma^\mathrm{fid}$.
 
Total cross sections, $\sigma^\mathrm{tot}$, are calculated by dividing $\sigma^\mathrm{fid}$ by the acceptance $A$. This correction is calculated
with the \SHERPA\ signal samples by dividing the predicted fiducial cross section by the predicted total cross section at the particle level,
the latter derived without any kinematic cuts in the \W-boson case and only with an invariant mass window $66<m_{\ell\ell}<116$\,\gev in the \Z-boson case.
The acceptance $A$ includes theory uncertainties (Section~\ref{sec:unc}), which are then propagated into $\sigma^\mathrm{tot}$.
 
The impact of the uncertainties in the measured fiducial cross sections arising from different sources is summarised in Table~\ref{tab:error_breakdown}. The uncertainties are grouped into distinct categories.
To estimate the impact associated with each category, the nuisance parameters belonging to the category
are fixed at the best-fit values and set as constants. Then a fitting process is performed again, resulting in a decreased uncertainty compared
to the nominal uncertainty values. The difference in quadrature between the new and the nominal uncertainty is the impact of this category.
The luminosity uncertainty values vary slightly from one channel to another due to different background contributions in each channel.
The statistical impact is obtained by repeating the fit after having fixed all nuisance parameters to their fitted values.
 
\begin{table}[tb]
\centering
\caption{Observed impact (in \%) of the different sources of uncertainty on the measured fiducial cross sections. Flavour-tagging uncertainties and top-quark modelling uncertainties are taken from Ref.~\cite{TOPQ-2023-21}.}
\label{tab:error_breakdown}

\scalebox{0.7}{
\begin{tabular}{c|c|c|c|c|c|c|c}
\hline
Category                                     & $\sigma(Z\to ee)$ & $\sigma(Z\to \mu\mu)$ & $\sigma(Z\to \ell\ell)$ & $\sigma(W^{-}\rightarrow e^-\bar{\nu})$ & $\sigma(W^{+}\rightarrow e^+\nu)$ & $\sigma(W^{-}\rightarrow \mu^-\bar{\nu})$ & $\sigma(W^{+}\rightarrow \mu^+\nu)$ \\ \hline
Luminosity                                   & 2.2               & 2.2                   & 2.2                     & 2.5                                     & 2.5                               & 2.5                                       & 2.4                                 \\
Pile-up                                      & 1.2               & 0.3                   & 0.8                     & 1.1                                     & 1.1                               & 0.3                                       & 0.4                                 \\
MC statistics                                & $<0.2$            & $<0.2$                & $<0.2$                  & $<0.2$                                  & 0.4                               & $<0.2$                                    & 0.4                                 \\
Lepton trigger                               & 0.2               & 0.4                   & 0.2                     & 1.2                                     & 1.3                               & 1.0                                       & 1.0                                 \\
Electron reconstruction                      & 1.4               & --                    & 0.9                     & 0.7                                     & 0.8                               & --                                        & --                                  \\
Muon reconstruction                          & --                & 2.1                   & 1.4                     & --                                      & --                                & 1.0                                       & 1.0                                 \\
Multi-jet                                    & --                & --                    & --                      & 2.9                                     & 2.4                               & 1.3                                       & 1.1                                 \\
Other background modelling                   & $<0.2$            & $<0.2$                & $<0.2$                  & $<0.2$                                  & $<0.2$                            & 0.5                                       & 0.4                                 \\
Jet energy scale                             & --                & --                    & --                      & 1.4                                     & 1.4                               & 1.3                                       & 1.4                                 \\
Jet energy resolution                        & --                & --                    & --                      & $<0.2$                                  & 0.3                               & 0.2                                       & 0.2                                 \\
NNJVT                            & --                & --                    & --                      & 1.6                                     & 1.5                               & 1.3                                       & 1.3                                 \\
$E_\mathrm{T}^\mathrm{miss}$ track soft term & --                & --                    & --                      & $<0.2$                                  & 0.4                               & $<0.2$                                    & $<0.2$                              \\
PDF                                          & 0.2               & 0.2                   & $<0.2$                  & 0.8                                     & 0.8                               & 0.6                                       & 0.5                                 \\
QCD scale (ME and PS)                        & 0.6               & $<0.2$                & 0.3                     & 1.3                                     & 1.2                               & 0.6                                       & 0.6                                 \\
Flavour tagging                              & --                & --                    & --                      & --                                      & --                                & --                                        & --                                  \\
$t\bar{t}$ modelling                         & --                & --                    & --                      & --                                      & --                                & --                                        & --                                  \\ \hline
Total systematic impact [\%]                      & 3.0               & 3.1                   & 2.7                     & 5.0                                     & 4.5                               & 3.8                                       & 3.6                                 \\
Statistical impact [\%]                       & 0.04              & 0.03                  & 0.02                    & 0.02                                    & 0.01                              & 0.01                                      & 0.01                                \\ \hline
\end{tabular}
}

\scalebox{0.7}{
\begin{tabular}{c|c|c|c|c|c|c}
\hline
Category                                     & $\sigma(W^{-}\rightarrow \ell^-\bar{\nu})$ & $\sigma(W^{+}\rightarrow \ell^+\nu)$ & $\sigma(W^{\pm}\rightarrow \ell\nu)$ & $R_{W^{+}/W^{-}}$ & $R_{W^{\pm}/Z}$ & $R_{t\bar{t}/W^{\pm}}$ \\\hline
Luminosity                                   & 2.5                                        & 2.4                                  & 2.4                                  & $<0.2$            & 0.3             & $<0.2$                 \\
Pile-up                                      & 0.5                                        & 0.7                                  & 0.6                                  & $<0.2$            & $<0.2$          & $<0.2$                 \\
MC statistics                                & $<0.2$                                     & 0.2                                  & $<0.2$                               & $<0.2$            & $<0.2$          & $<0.2$                 \\
Lepton trigger                               & 1.0                                        & 0.9                                  & 0.9                                  & $<0.2$            & 0.7             & 0.8                    \\
Electron reconstruction                      & 0.4                                        & 0.5                                  & 0.4                                  & $<0.2$            & 0.5             & 0.4                    \\
Muon reconstruction                          & 0.6                                        & 0.6                                  & 0.6                                  & 0.2               & 0.8             & 0.6                    \\
Multi-jet                                    & 1.2                                        & 1.2                                  & 1.2                                  & 1.6               & 1.1             & 1.0                    \\
Other background modelling                   & 0.4                                        & 0.4                                  & 0.4                                  & $<0.2$            & 0.3             & 0.9                    \\
Jet energy scale                             & 1.3                                        & 1.3                                  & 1.3                                  & $<0.2$            & 1.3             & 1.3                    \\
Jet energy resolution                        & $<0.2$                                     & 0.2                                  & $<0.2$                               & $<0.2$            & $<0.2$          & $<0.2$                 \\
NNJVT                            & 1.4                                        & 1.3                                  & 1.3                                  & $<0.2$            & 1.3             & $<0.2$                 \\
$E_\mathrm{T}^\mathrm{miss}$ track soft term & $<0.2$                                     & 0.3                                  & 0.3                                  & $<0.2$            & 0.3             & 0.3                    \\
PDF                                          & 0.5                                        & 0.5                                  & 0.3                                  & 0.5               & 0.2             & 0.4                    \\
QCD scale (ME and PS)                        & 0.8                                        & 0.7                                  & 0.6                                  & $<0.2$            & 0.7             & 0.7                    \\
Flavour tagging                              & --                                         & --                                   & --                                   & --                & --              & $<0.2$                 \\
$t\bar{t}$ modelling                         & --                                         & --                                   & --                                   & --                & --              & 1.1                    \\ \hline
Total systematic impact [\%]                        & 3.7                                        & 3.5                                  & 3.5                                  & 1.7               & 2.4             & 2.5                    \\
Statistical impact [\%]                        & 0.01                                       & 0.01                                 & 0.01                                 & 0.01              & 0.02            & 0.32                   \\
\hline
\end{tabular}
}


\end{table}

The comparison of data and predictions before and after fits in all regions is shown in Figure~\ref{fig:pre_post_fit}. Good agreement is observed in the single lepton and same flavour di-lepton regions while in the $e\mu$ regions the data event yields are slightly lower than the predictions.
The measured fiducial and total cross-section results and the corresponding acceptance with their respective uncertainties are summarised in Table~\ref{tab:lh_results}.
The \Z-boson cross sections (\sZtoee, \sZtomm, and \sZtoll) are obtained through individual fits to the two same flavour dilepton regions. The \W-boson cross sections and the fiducial ratios
(\RWpm\ and \RWZ) are fitted  using the four single lepton and the two same flavour dilepton regions. Several fits are performed, using the same inputs
but changing the parameters of interest each time. The \Wptoenu, \Wptomunu, \Wmtoenu, and \Wmtomunu\ cross sections are obtained
simultaneously. The \Wp- and  \Wm-boson cross sections are also derived in a simultaneous fit. The \Wpm-boson cross section, the \RWpm\  ratio, and
the \RWZ\ ratio are obtained in separate fits.
The \tt\ to \W-boson cross-section ratios \RWtt, \RWptt, and \RWmtt\ are derived in separate fits to the four \W-boson channels, the two \Z-boson channels,
and the two \tt\ \emu\ channels from Ref.~\cite{TOPQ-2023-21}.
 
\begin{figure}[tb]
\subfloat[]{
\includegraphics[width=0.51\textwidth]{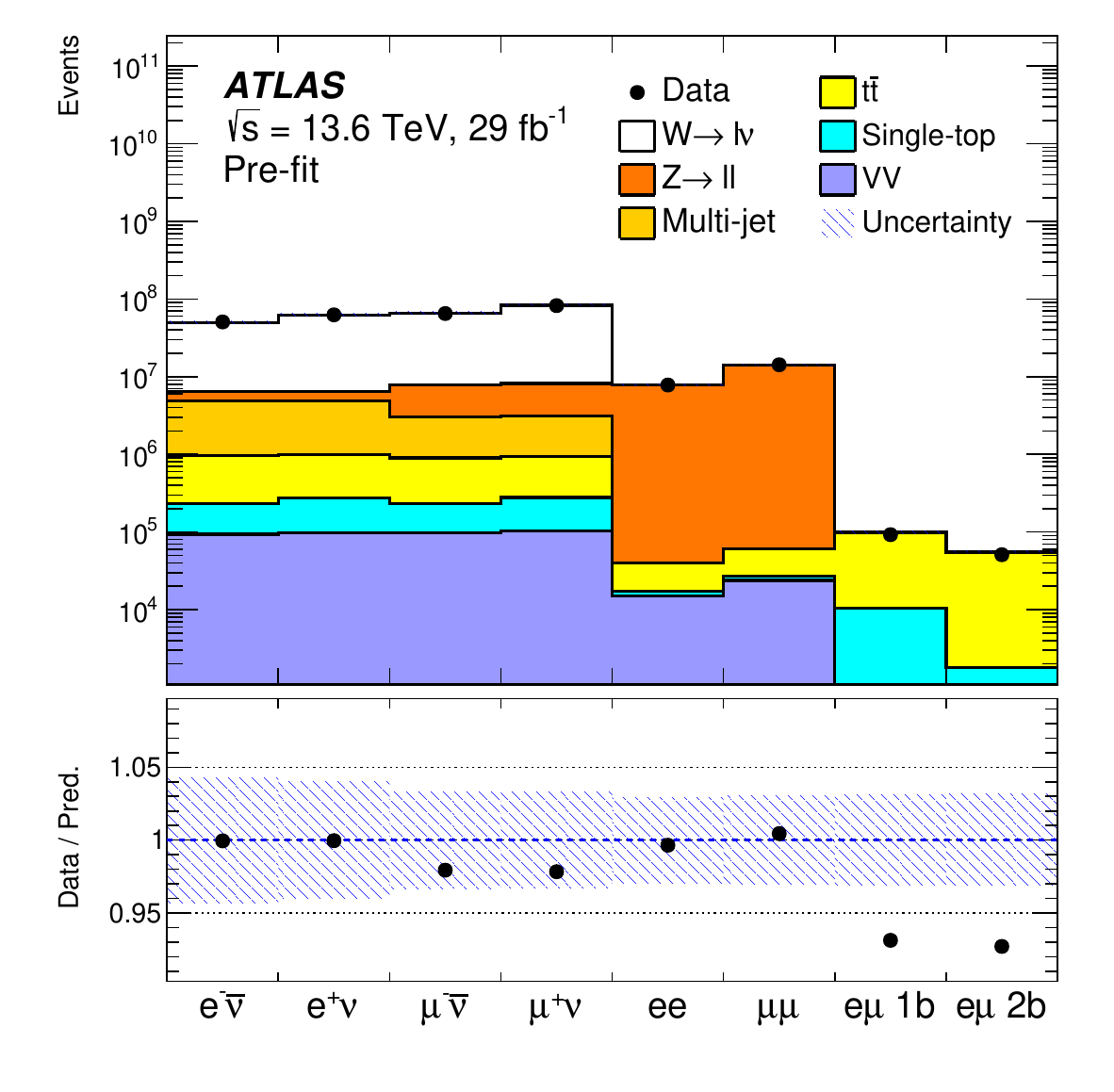}
}
\subfloat[]{
\includegraphics[width=0.51\textwidth]{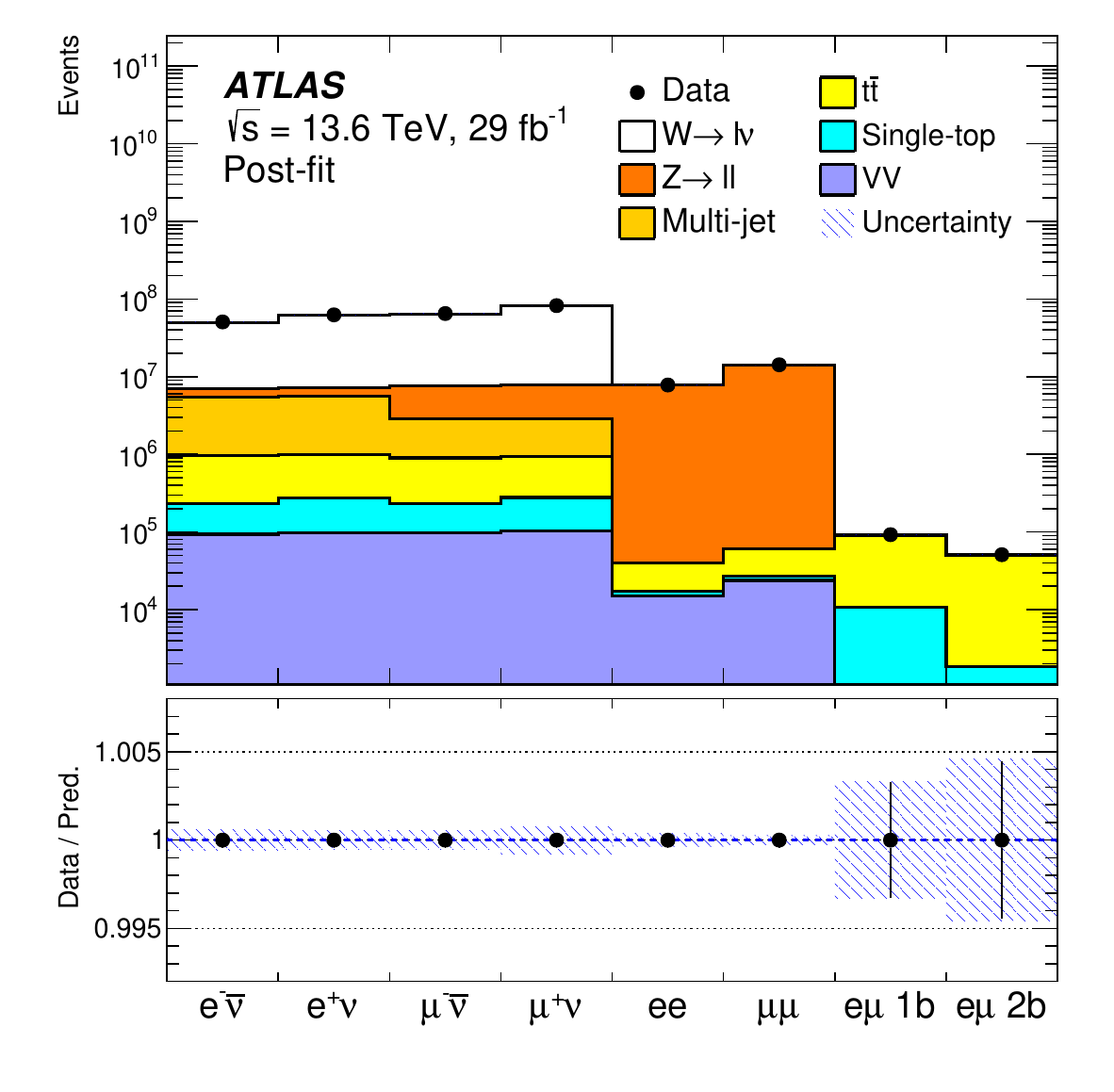}
}
\caption{Comparison of the number of data events in each channel (dots) with the predictions (stacked histograms) shown (a) before and (b) after the fits. The dashed error band in the pre-fit figure gives the total systematic uncertainty before the fit, while in the post-fit figure, it represents the statistical uncertainty derived from the fit.}
\label{fig:pre_post_fit}
\end{figure}
 
\begin{table}[tb]
\centering
\caption{The measured cross sections using the profile likelihood method. The quoted uncertainty corresponds to the total uncertainty (the statistical uncertainty is negligibly small). Rounding has been applied to all quoted numbers.}
\renewcommand{\arraystretch}{1.2}
\label{tab:lh_results}

\newcolumntype{P}[1]{>{\centering\arraybackslash}p{#1}}
\begin{tabular}{l|ccc}
\hline
Channel & $\sigma^\mathrm{fid}\pm \delta\sigma_\mathrm{stat\,\oplus\, syst}$ [pb] & Acceptance $A$ & $\sigma^\mathrm{tot}\pm \delta\sigma_\mathrm{stat\,\oplus\, syst}$ [pb] \\
\hline
$Z\to e^{+}e^{-}$         & $740\pm{22}$ & $0.374 \pm 0.011$ & $1981\pm{82}$ \\
$Z\to \mu^{+}\mu^{-}$ & $747\pm{23}$ & $0.374 \pm 0.011$ & $1997\pm{82}$ \\
$Z\to \ell^{+}\ell^{-}$    & $744\pm{20}$ & $0.374 \pm 0.011$ & $1989\pm{77}$ \\\hline
$W^{-}\to e^{-}\bar{\nu}$     & $3380\pm{170}$ & $0.381 \pm 0.009$ & $8880\pm{490}$ \\
$W^{-}\to \mu^{-}\bar{\nu}$ & $3310\pm{130}$ & $0.381 \pm 0.009$ & $8680\pm{390}$ \\
$W^{-}\to \ell^{-}\bar{\nu}$  & $3310\pm{120}$ & $0.381 \pm 0.009$ & $8690\pm{390}$ \\\hline
$W^{+}\to e^{+}\nu$     & $4350\pm{200}$ & $0.366 \pm 0.009$ & $11880\pm{620}$ \\
$W^{+}\to \mu^{+}\nu$ & $4240\pm{160}$ & $0.365 \pm 0.010$ & $11620\pm{530}$ \\
$W^{+}\to \ell^{+}\nu$  & $4250\pm{150}$ & $0.366 \pm 0.009$ & $11620\pm{520}$ \\\hline
$W^{\pm}\to \ell^{\pm}\nu$ & $7560\pm{270}$ & $0.372 \pm 0.009$ & $20310\pm{890}$ \\\hline
\end{tabular}
\begin{tabular}{l|c}	\hline
Ratio & $R \pm\delta R_\mathrm{stat\,\oplus\, syst}$ \\\hline
$W^{+}/W^{-}$ & $1.286\pm0.022$ \\
$W^{\pm}/Z$   & $10.17\pm0.25$ \\
$t\bar{t}/W^{-}$ & $0.256\pm0.008$ \\
$t\bar{t}/W^{+}$ & $0.199\pm{0.006}$ \\
$t\bar{t}/W^\pm$ & $0.112\pm{0.003}$ \\\hline
\end{tabular}


\end{table}
 
Figure~\ref{fig:DataTheory1} compares the measured results for the \Wm, \Wp, and \Z\ boson fiducial cross sections and \RWZ\, \RWpm, and $t\bar{t}/W$ ratios to the Standard-Model (SM) theory predictions with different PDF sets: PDF4LHC21~\cite{Ball:2022hsh}, CT18, CT18A~\cite{Hou:2019efy}, MSHT20~\cite{Cridge:2021qfd,Bailey:2020ooq,Cridge:2021pxm}, NNPDF4.0~\cite{Ball:2021leu}, ABMP16~\cite{Alekhin:2017kpj}, and ATLASpdf21~\cite{STDM-2020-32}.
The predictions are calculated at NNLO+NNLL QCD and NLO EW accuracy using  \textsc{DYTurbo-1.3.1}~\cite{Camarda:2019zyx,Camarda:2021ict,Camarda:2021jsw,Camarda:2023dqn} and \textsc{ReneSANCe-1.3.3}~\cite{Sadykov:2020any,Bondarenko:2022mbi}, and combining them with  an additive prescription, as used in Ref.~\cite{STDM-2012-20}.
The nominal $t\bar{t}$ predictions for all PDFs are calculated at NNLO in QCD including the resummation of NNLL soft-gluon terms calculated using \TOPpp[2.0]~\cite{Beneke:2011mq,Cacciari:2011hy,Baernreuther:2012ws,Czakon:2012zr,Czakon:2012pz,Czakon:2013goa,Czakon:2011xx} based on $m_t = 172.5$\,GeV.
Theoretical uncertainties arising from normalisation and factorisation scale variations for missing higher orders, the variation of the strong coupling constant $\alpha_s$, and variations of the input PDFs are evaluated. The latter source dominates in most cases as indicated with the inner error bars.
Scale variations are taken as uncorrelated between different bosons and between bosons and \ttbar, but the PDF variations are treated as correlated.
 
For the \Wm, \Wp, and \Z\ boson fiducial cross sections and their ratios, an overall good agreement is observed, while the $t\bar{t}/W$ ratio results are slightly lower than the predictions for most of the PDFs considered. This is consistent with the results of the Run~3 \tt\ cross-section measurement~\cite{TOPQ-2023-21},
where the measured \tt\ cross-section is measured to be lower than the predicted value. The predictions based on the PDF4LHC21 set with $m_t = 171.5$\,GeV and $m_t = 173.5$\,GeV are also included in Figure~\ref{fig:DataTheory1}.
 
\begin{figure}[tb]
\begin{center}
\subfloat[]{
\includegraphics[width=0.75\textwidth]{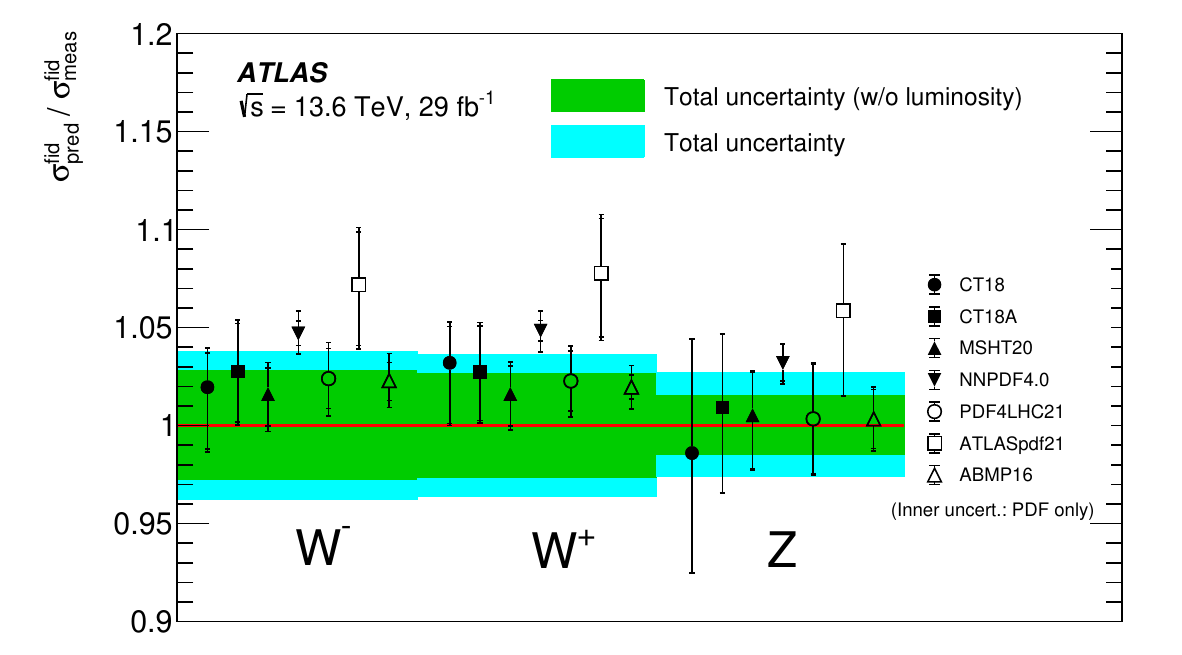}
}\\
\subfloat[]{
\includegraphics[width=0.42\textwidth]{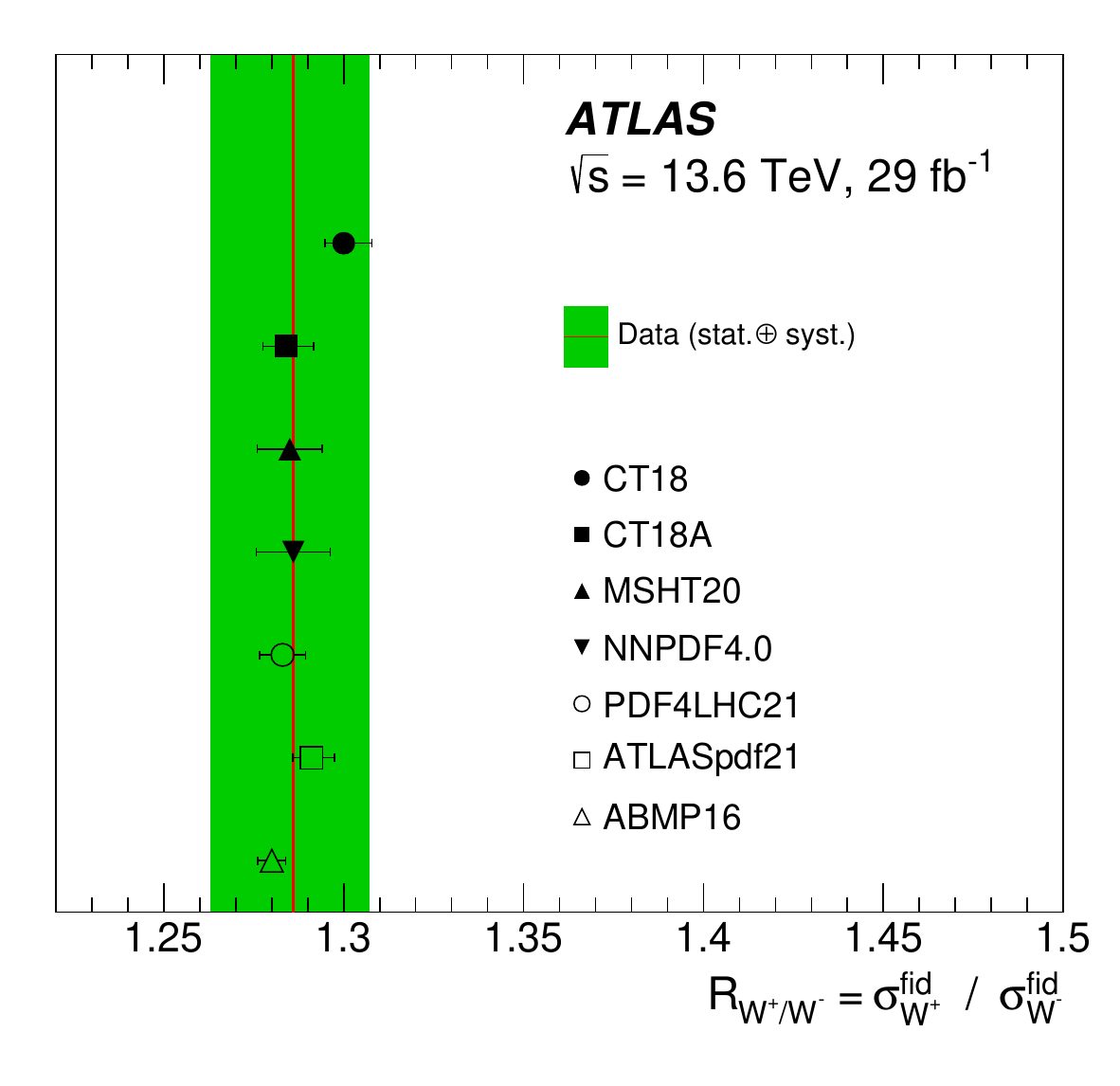}
}
\subfloat[]{
\includegraphics[width=0.42\textwidth]{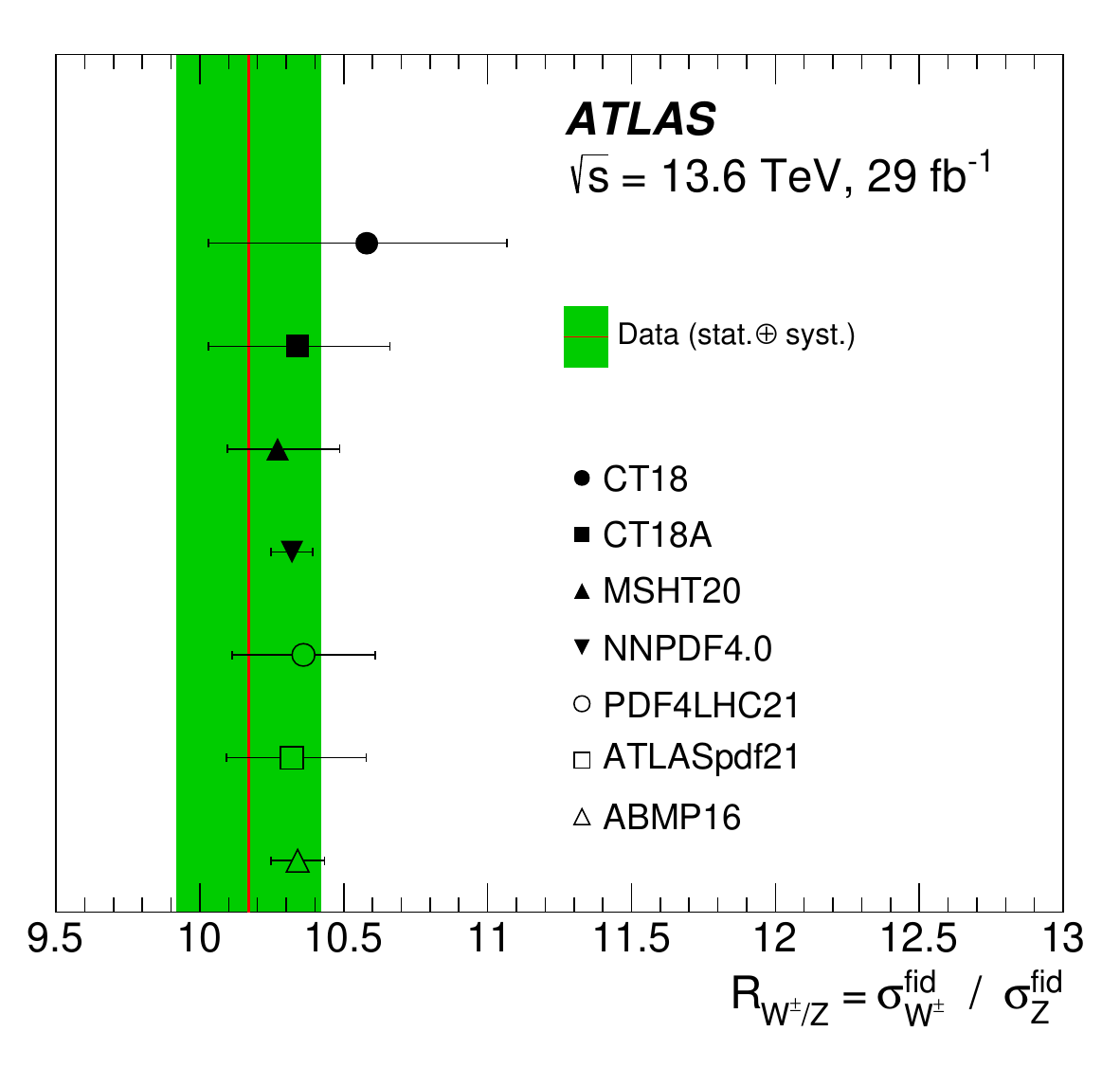}
}
\end{center}
\caption{The ratio of the predictions obtained with different PDF sets and the measured fiducial cross sections for (a) $W^-$, $\W^+$, and  $Z$ bosons, (b) and (c) their ratios $R_{W^+/W^-}$ and $R_{W^\pm/Z}$, (d), (e) and (f) ratios of $t\bar{t}$ over $W$-boson fiducial cross sections $R_{t\bar{t}/W^\pm}$, $R_{t\bar{t}/W^+}$, and $R_{t\bar{t}/W^-}$.
The outer (inner) band in (a) corresponds to the total uncertainty including (excluding) the luminosity uncertainty.
The vertical band in the other plots shows the total (systematic and statistical) uncertainty in the data.
The error bars on the predictions correspond to the theory uncertainties with the inner error bars (where available) representing the contributions from the PDF uncertainty. }
\label{fig:DataTheory1}
\end{figure}
\begin{figure}[!htb]\ContinuedFloat
\begin{center}
\subfloat[]{
\includegraphics[width=0.42\textwidth]{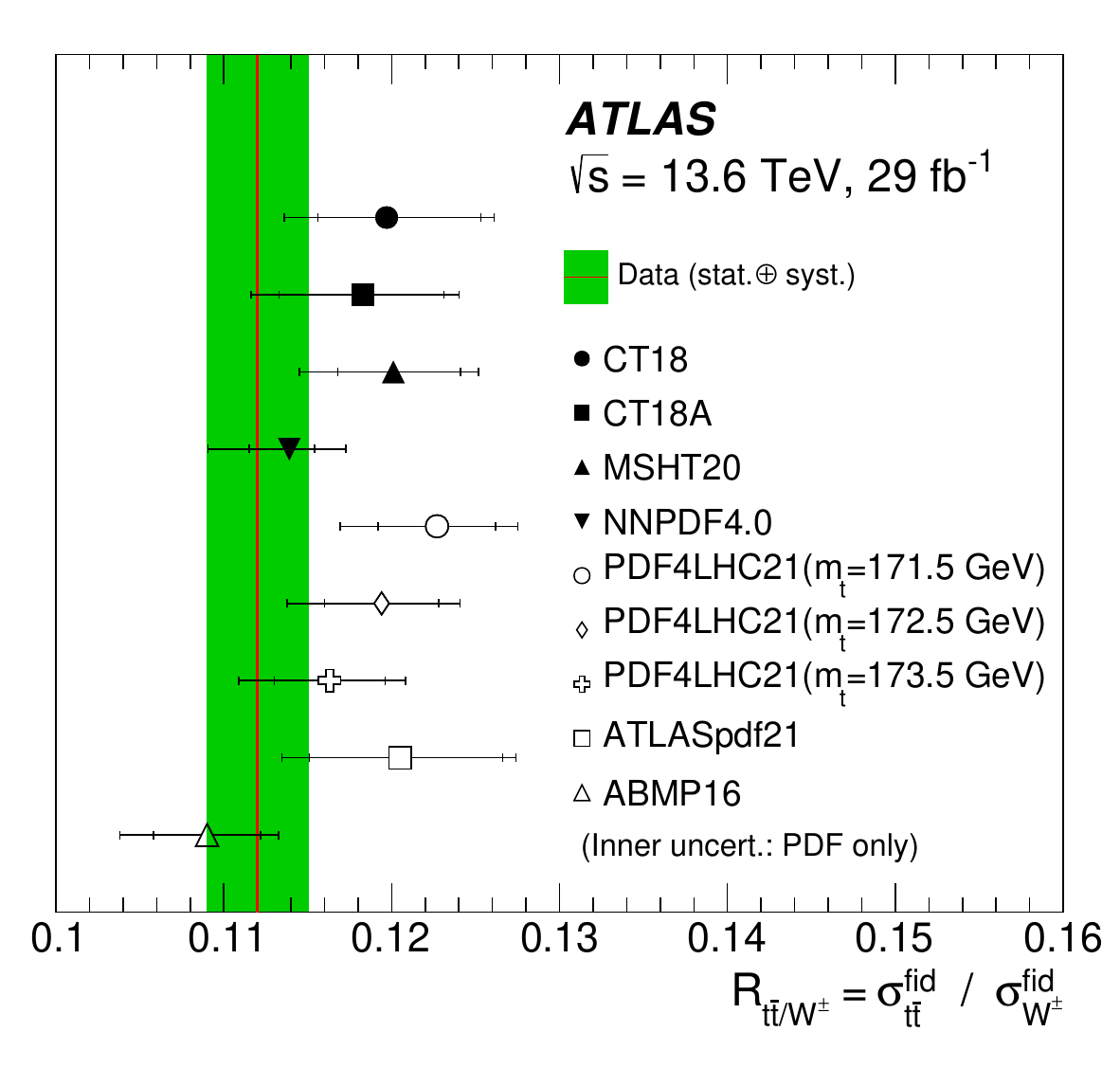}
}
\subfloat[]{
\includegraphics[width=0.42\textwidth]{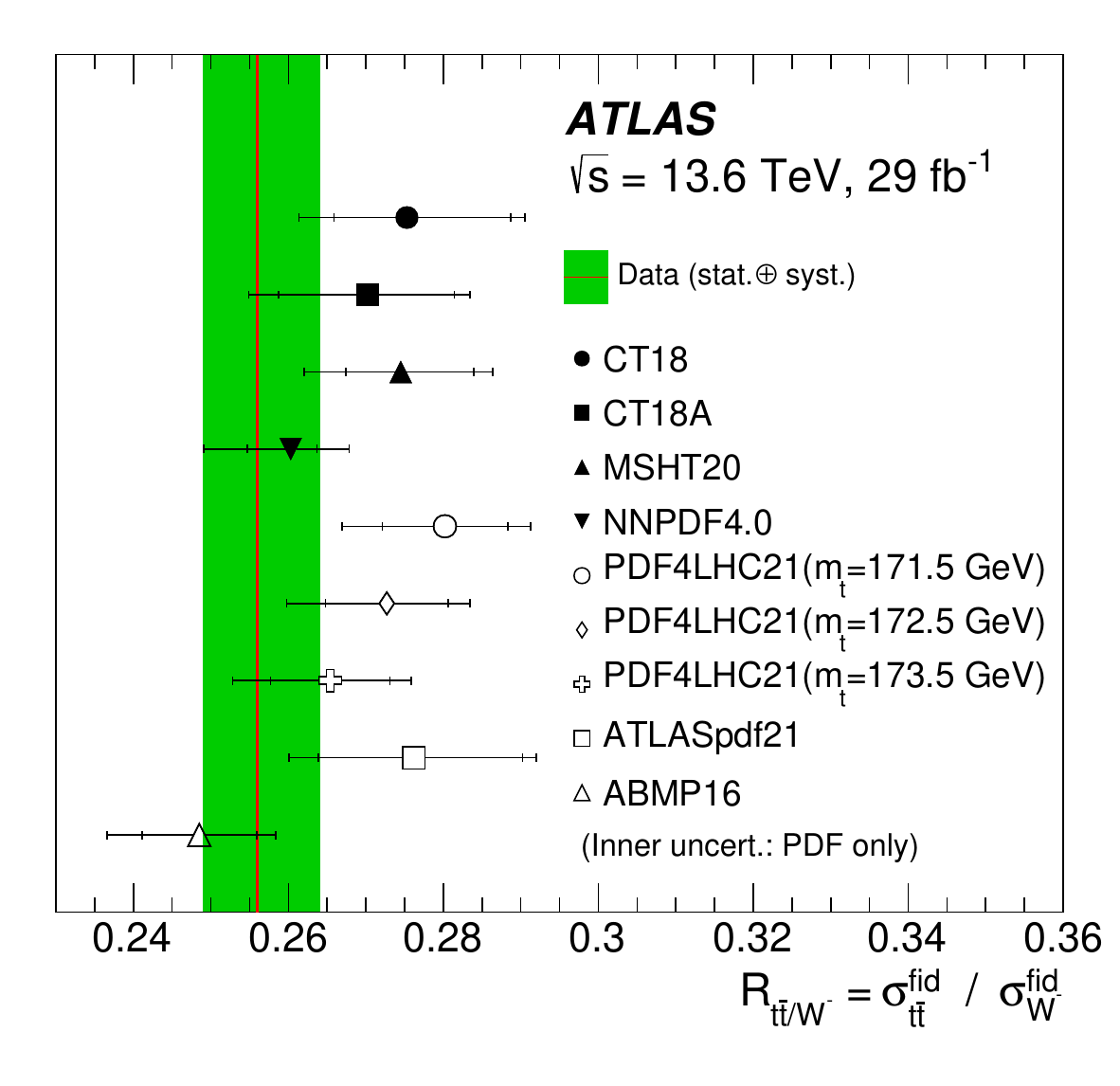}
}\\
\subfloat[]{
\includegraphics[width=0.42\textwidth]{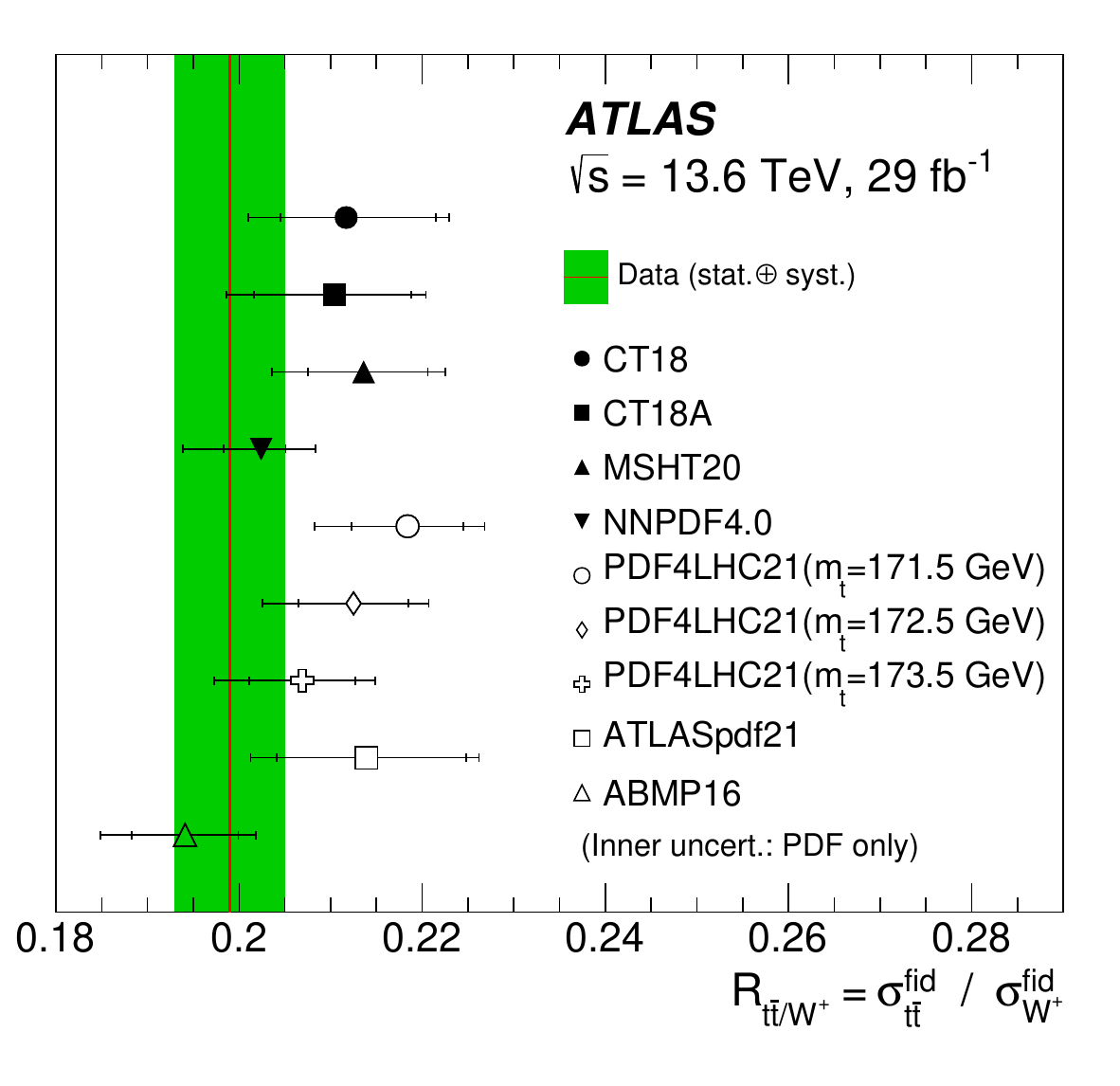}
}
\end{center}
\caption{The ratio of the predictions obtained with different PDF sets and the measured fiducial cross sections for (a) $W^-$, $\W^+$, and  $Z$ bosons, (b) and (c) their ratios $R_{W^+/W^-}$ and $R_{W^\pm/Z}$, (d), (e) and (f) ratios of $t\bar{t}$ over $W$-boson fiducial cross sections $R_{t\bar{t}/W^\pm}$, $R_{t\bar{t}/W^+}$, and $R_{t\bar{t}/W^-}$.
The outer (inner) band in (a) corresponds to the total uncertainty including (excluding) the luminosity uncertainty.
The vertical band in the other plots shows the total (systematic and statistical) uncertainty in the data.
The error bars on the predictions correspond to the theory uncertainties with the inner error bars (where available) representing the contributions from the PDF uncertainty. }
\end{figure}
 
The measured total cross section is shown as a function of $\sqrt{s}$, together compared with the CT14NNLO predictions~\cite{Dulat:2015mca} in Figure~\ref{fig:sigma_vs_sqrt}. Good agreement is observed between data and prediction.
 
\begin{figure}[tb]
\includegraphics[width=0.9\textwidth]{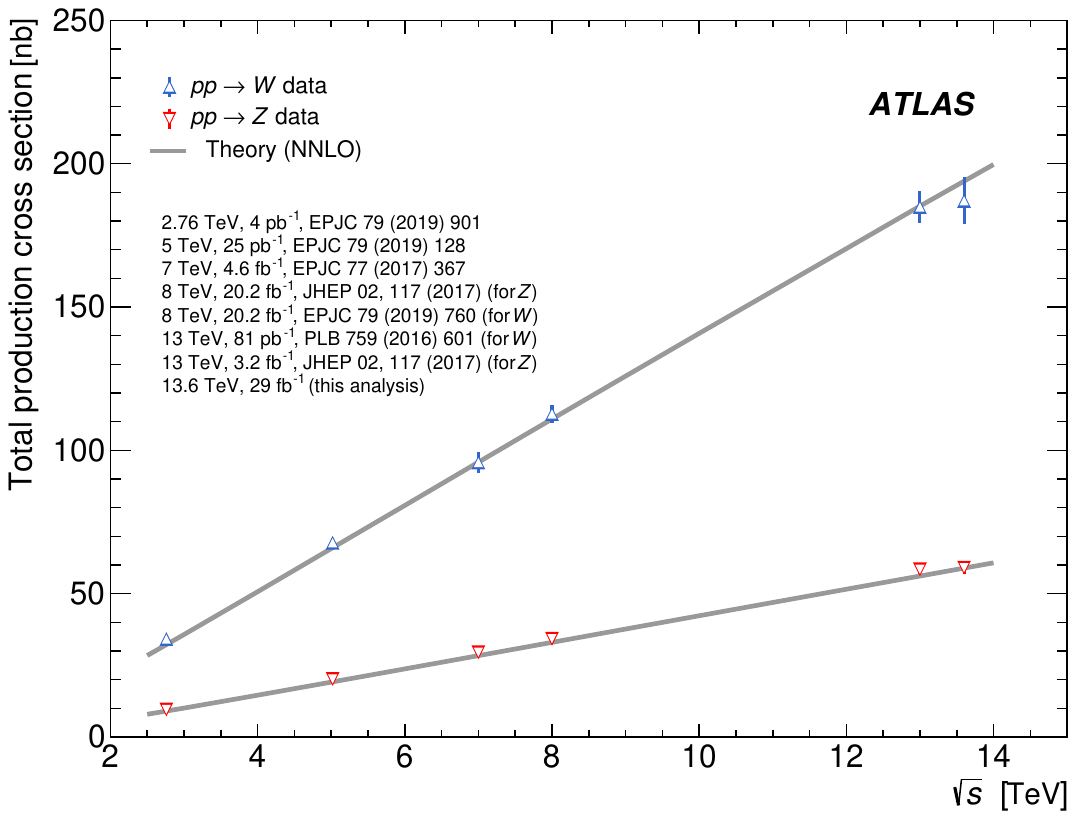}
\caption{The measured $W$ and $Z$ boson total production cross sections in the leptonic decay channel at different values of centre-of-mass energy.
For comparison, the central values of the NNLO predictions based on the CT14NNLO PDF set are included.}
\label{fig:sigma_vs_sqrt}
\end{figure}


\clearpage
\section{Conclusion}
 
A measurement of the inclusive \W-  and \Z-boson production cross sections with decays into final states with electrons or muons, their ratios, and the ratios
of \tt\ to \W-boson fiducial cross sections were performed using LHC $pp$ collision data corresponding to 29~\ifb\ collected by the ATLAS experiment at a new
centre-of-mass energy of 13.6~\TeV. The cross sections are measured for \Z-boson production in a fiducial phase space corresponding to $\pt^\ell>27$\,GeV,
$|\eta_\ell|<2.5$, and $66 <\mll<116~\gev$ and for $W$-boson production in a fiducial region corresponding to $\pt^\ell>27$~\GeV, $|\eta_\ell|<2.5$,
$\mtw>50$~\GeV, and $\met>25$~\GeV.
The measured fiducial cross-section values for $W^+$, $W^-$, and $Z$ boson productions are $4250\pm 150$\,pb, $3310±\pm 120$\,pb, and $744\pm 20$\,pb,
respectively, where the uncertainty corresponds to the total uncertainty, including that arising from the luminosity, which amounts to about 2.2\%. The measured values for the ratios of fiducial cross sections are: $R_{W^+/W^-}=1.286\pm0.022$,
$R_{W^\pm/Z}=10.17\pm0.25$, and $R_{t\bar{t}/W^{\pm}}=0.112\pm0.003$, where the uncertainty is the total uncertainty.
The total cross sections are also measured in the mass range $66~\gev<m_{\ell\ell}<116~\gev$ for \Z-boson production
and in the full phase space for \W-boson production. Measurements are performed on \sWp, \sWm, \sWpm, \sZ, \RWpm, \RWZ, \RWtt, \RWptt, and \RWmtt.
The measured \W- and \Z-boson cross sections are in good agreement with the SM predictions whereas the \tt\ over \W-boson fiducial cross-section ratios are slightly overestimated by some of the theoretical predictions.


\section*{Acknowledgements}


We thank CERN for the very successful operation of the LHC and its injectors, as well as the support staff at
CERN and at our institutions worldwide without whom ATLAS could not be operated efficiently.
 
The crucial computing support from all WLCG partners is acknowledged gratefully, in particular from CERN, the ATLAS Tier-1 facilities at TRIUMF/SFU (Canada), NDGF (Denmark, Norway, Sweden), CC-IN2P3 (France), KIT/GridKA (Germany), INFN-CNAF (Italy), NL-T1 (Netherlands), PIC (Spain), RAL (UK) and BNL (USA), the Tier-2 facilities worldwide and large non-WLCG resource providers. Major contributors of computing resources are listed in Ref.~\cite{ATL-SOFT-PUB-2023-001}.
 
We gratefully acknowledge the support of ANPCyT, Argentina; YerPhI, Armenia; ARC, Australia; BMWFW and FWF, Austria; ANAS, Azerbaijan; CNPq and FAPESP, Brazil; NSERC, NRC and CFI, Canada; CERN; ANID, Chile; CAS, MOST and NSFC, China; Minciencias, Colombia; MEYS CR, Czech Republic; DNRF and DNSRC, Denmark; IN2P3-CNRS and CEA-DRF/IRFU, France; SRNSFG, Georgia; BMBF, HGF and MPG, Germany; GSRI, Greece; RGC and Hong Kong SAR, China; ISF and Benoziyo Center, Israel; INFN, Italy; MEXT and JSPS, Japan; CNRST, Morocco; NWO, Netherlands; RCN, Norway; MNiSW, Poland; FCT, Portugal; MNE/IFA, Romania; MESTD, Serbia; MSSR, Slovakia; ARRS and MIZ\v{S}, Slovenia; DSI/NRF, South Africa; MICINN, Spain; SRC and Wallenberg Foundation, Sweden; SERI, SNSF and Cantons of Bern and Geneva, Switzerland; MOST, Taipei; TENMAK, T\"urkiye; STFC, United Kingdom; DOE and NSF, United States of America.
 
Individual groups and members have received support from BCKDF, CANARIE, CRC and DRAC, Canada; PRIMUS 21/SCI/017, CERN-CZ and FORTE, Czech Republic; COST, ERC, ERDF, Horizon 2020, ICSC-NextGenerationEU and Marie Sk{\l}odowska-Curie Actions, European Union; Investissements d'Avenir Labex, Investissements d'Avenir Idex and ANR, France; DFG and AvH Foundation, Germany; Herakleitos, Thales and Aristeia programmes co-financed by EU-ESF and the Greek NSRF, Greece; BSF-NSF and MINERVA, Israel; Norwegian Financial Mechanism 2014-2021, Norway; NCN and NAWA, Poland; La Caixa Banking Foundation, CERCA Programme Generalitat de Catalunya and PROMETEO and GenT Programmes Generalitat Valenciana, Spain; G\"{o}ran Gustafssons Stiftelse, Sweden; The Royal Society and Leverhulme Trust, United Kingdom.
 
In addition, individual members wish to acknowledge support from CERN: European Organization for Nuclear Research (CERN PJAS); Chile: Agencia Nacional de Investigaci\'on y Desarrollo (FONDECYT 1190886, FONDECYT 1210400, FONDECYT 1230812, FONDECYT 1230987); China: National Natural Science Foundation of China (NSFC - 12175119, NSFC 12275265, NSFC-12075060); Czech Republic: Czech Science Foundation (GACR - 24-11373S), Ministry of Education Youth and Sports (FORTE CZ.02.01.01/00/22\_008/0004632), PRIMUS Research Programme (PRIMUS/21/SCI/017); European Union: European Research Council (ERC - 948254, ERC 101089007), Horizon 2020 Framework Programme (MUCCA - CHIST-ERA-19-XAI-00), European Union, Future Artificial Intelligence Research (FAIR-NextGenerationEU PE00000013), Italian Center for High Performance Computing, Big Data and Quantum Computing (ICSC, NextGenerationEU); France: Agence Nationale de la Recherche (ANR-20-CE31-0013, ANR-21-CE31-0013, ANR-21-CE31-0022), Investissements d'Avenir Labex (ANR-11-LABX-0012); Germany: Baden-Württemberg Stiftung (BW Stiftung-Postdoc Eliteprogramme), Deutsche Forschungsgemeinschaft (DFG - 469666862, DFG - CR 312/5-2); Italy: Istituto Nazionale di Fisica Nucleare (ICSC, NextGenerationEU); Japan: Japan Society for the Promotion of Science (JSPS KAKENHI JP21H05085, JSPS KAKENHI JP22H01227, JSPS KAKENHI JP22H04944, JSPS KAKENHI JP22KK0227); Netherlands: Netherlands Organisation for Scientific Research (NWO Veni 2020 - VI.Veni.202.179); Norway: Research Council of Norway (RCN-314472); Poland: Polish National Agency for Academic Exchange (PPN/PPO/2020/1/00002/U/00001), Polish National Science Centre (NCN 2021/42/E/ST2/00350, NCN OPUS nr 2022/47/B/ST2/03059, NCN UMO-2019/34/E/ST2/00393, UMO-2020/37/B/ST2/01043, UMO-2021/40/C/ST2/00187, UMO-2022/47/O/ST2/00148, UMO-2023/49/B/ST2/04085); Slovenia: Slovenian Research Agency (ARIS grant J1-3010); Spain: Generalitat Valenciana (Artemisa, FEDER, IDIFEDER/2018/048), Ministry of Science and Innovation (MCIN \& NextGenEU -PCI2022-135018-2, MICIN \& FEDER -PID2021-125273NB, RYC2019-028510-I, RYC2020-030254-I, RYC2021-031273-I, RYC2022-038164-I), PROMETEO and GenT Programmes Generalitat Valenciana (CIDEGENT/2019/023, CIDEGENT/2019/027); Sweden: Swedish Research Council (Swedish Research Council 2023-04654, VR 2018-00482, VR 2022-03845, VR 2022-04683, VR 2023-03403, VR grant 2021-03651), Knut and Alice Wallenberg Foundation (KAW 2018.0157, KAW 2018.0458, KAW 2019.0447, KAW 2022.0358); Switzerland: Swiss National Science Foundation (SNSF - PCEFP2\_194658); United Kingdom: Leverhulme Trust (Leverhulme Trust RPG-2020-004), Royal Society (NIF-R1-231091); United States of America: Neubauer Family Foundation.


\printbibliography
\clearpage
\clearpage
\input{atlas_authlist}

\end{document}

%% file: atlas_authlist.tex
 
\begin{flushleft}
\hypersetup{urlcolor=black}
{\Large The ATLAS Collaboration}

\bigskip

\AtlasOrcid[0000-0002-6665-4934]{G.~Aad}$^\textrm{\scriptsize 103}$,
\AtlasOrcid[0000-0001-7616-1554]{E.~Aakvaag}$^\textrm{\scriptsize 16}$,
\AtlasOrcid[0000-0002-5888-2734]{B.~Abbott}$^\textrm{\scriptsize 121}$,
\AtlasOrcid[0000-0002-0287-5869]{S.~Abdelhameed}$^\textrm{\scriptsize 117a}$,
\AtlasOrcid[0000-0002-1002-1652]{K.~Abeling}$^\textrm{\scriptsize 55}$,
\AtlasOrcid[0000-0001-5763-2760]{N.J.~Abicht}$^\textrm{\scriptsize 49}$,
\AtlasOrcid[0000-0002-8496-9294]{S.H.~Abidi}$^\textrm{\scriptsize 29}$,
\AtlasOrcid[0009-0003-6578-220X]{M.~Aboelela}$^\textrm{\scriptsize 44}$,
\AtlasOrcid[0000-0002-9987-2292]{A.~Aboulhorma}$^\textrm{\scriptsize 35e}$,
\AtlasOrcid[0000-0001-5329-6640]{H.~Abramowicz}$^\textrm{\scriptsize 152}$
% ...
\bigskip
\\

% ...
$^{16}$Department for Physics and Technology, University of Bergen, Bergen; Norway.\\
% ...
$^{29}$Physics Department, Brookhaven National Laboratory, Upton NY; United States of America.\\
% ...
$^{35}$$^{(a)}$Facult\'e des Sciences Ain Chock, R\'eseau Universitaire de Physique des Hautes Energies - Universit\'e Hassan II, Casablanca;$^{(b)}$Facult\'{e} des Sciences, Universit\'{e} Ibn-Tofail, K\'{e}nitra;$^{(c)}$Facult\'e des Sciences Semlalia, Universit\'e Cadi Ayyad, LPHEA-Marrakech;$^{(d)}$LPMR, Facult\'e des Sciences, Universit\'e Mohamed Premier, Oujda;$^{(e)}$Facult\'e des sciences, Universit\'e Mohammed V, Rabat;$^{(f)}$Institute of Applied Physics, Mohammed VI Polytechnic University, Ben Guerir; Morocco.\\
% ...
$^{44}$Physics Department, Southern Methodist University, Dallas TX; United States of America.\\
% ...
$^{49}$Fakult\"{a}t Physik , Technische Universit{\"a}t Dortmund, Dortmund; Germany.\\
% ...
$^{55}$II. Physikalisches Institut, Georg-August-Universit\"{a}t G\"ottingen, G\"ottingen; Germany.\\
% ...
$^{103}$CPPM, Aix-Marseille Universit\'e, CNRS/IN2P3, Marseille; France.\\
% ...
$^{117}$$^{(a)}$New York University Abu Dhabi, Abu Dhabi;$^{(b)}$United Arab Emirates University, Al Ain; United Arab Emirates.\\
% ...
$^{121}$Homer L. Dodge Department of Physics and Astronomy, University of Oklahoma, Norman OK; United States of America.\\
% ...
$^{152}$Raymond and Beverly Sackler School of Physics and Astronomy, Tel Aviv University, Tel Aviv; Israel.\\
% ...
$^{a}$ Also Affiliated with an institute covered by a cooperation agreement with CERN.\\
$^{b}$ Also at An-Najah National University, Nablus; Palestine.\\
$^{c}$ Also at Borough of Manhattan Community College, City University of New York, New York NY; United States of America.\\
$^{d}$ Also at Center for Interdisciplinary Research and Innovation (CIRI-AUTH), Thessaloniki; Greece.\\
$^{e}$ Also at Centro Studi e Ricerche Enrico Fermi; Italy.\\
$^{f}$ Also at CERN, Geneva; Switzerland.\\
% ...

\end{flushleft}
